\documentclass{IEEEtran}
\usepackage{amsmath,amssymb,amsfonts,bm}
\usepackage{graphicx,subfig}
\usepackage{cite}
\ifCLASSINFOpdf
\else
\fi
\begin{document}
%
\title{RFI Mitigation for One-bit UWB Radar Systems}
%
%
%

\author{Tianyi~Zhang,~
Jiaying~Ren,
Jian~Li,~\IEEEmembership{Fellow,~IEEE,}
Lam~H.~Nguyen
and~Petre~Stoica,~\IEEEmembership{Fellow,~IEEE,}
\thanks{This work was supported in part by the National Science Foundation under Grant 1704240, in part by the U.S. Army Research Laboratory and the U.S. Army Research Office under Grant W911NF-16-2-0223 and in part by the Swedish Research Council under VR grants 2017-04610 and 2016-06079 (corresponding author: Jian Li).}
\thanks{T. Zhang, J. Ren, and J. Li are with the department of Electrical and Computer Engineering, University of Florida, Gainesville, FL 32611, USA (e-mail: tianyi.zhang@ufl.edu, jiaying.ren@ufl.edu, li@dsp.ufl.edu).}
\thanks{L.H.Nguyen is with the U.S. Army Research Laboratory, Adelphi, MD 20783 USA (e-mail: lam.h.nguyen2.civ@mail.mil)}
\thanks{P. Stoica is with the department of Information Technology, Uppsala University, P. O. Box 337, SE-751 05 Uppsala, Sweden (e-mail: ps@it.uu.se).}
}

\maketitle

\begin{abstract}
Radio frequency interference (RFI) mitigation is critical to the proper operation of ultra-wideband (UWB) radar systems since RFI can severely degrade the radar imaging capability and target detection performance. In this paper, we address the RFI mitigation problem for one-bit UWB radar systems. A one-bit UWB system obtains its signed measurements via a low-cost and high rate sampling scheme, referred to as the Continuous Time Binary Value (CTBV) technology. This sampling strategy compares the signal to a known threshold varying with slow-time and therefore can be used to achieve a rather high sampling rate and quantization resolution with rather simple and affordable hardware. This paper establishes a proper data model for the RFI sources and proposes a novel RFI mitigation method for the one-bit UWB radar system that uses the CTBV sampling technique. Specifically, we first model the RFI sources as a sum of sinusoids with frequencies fixed during the coherent processing interval (CPI) and we exploit the sparsity of the RFI spectrum. We extend a majorization-minimization based 1bRELAX algorithm, referred to as 1bMMRELAX, to estimate the RFI source parameters from the signed measurements obtained by using the CTBV sampling strategy. We also devise a new fast frequency initialization method based on the Alternating Direction Method of Multipliers (ADMM) methodology for the extended 1bMMRELAX algorithm to significantly improve its computational efficiency. Moreover, an ADMM-based sparse method is introduced to recover the desired radar echoes using the estimated RFI parameters. Both simulated and experimental results are presented to demonstrate that our proposed algorithm outperforms the existing digital integration method, especially for severe RFI cases.
\end{abstract}


%
\IEEEpeerreviewmaketitle

\section{Introduction}
%
%
%
%
\IEEEPARstart{U}{ltra-wideband} (UWB) radar has been used in a wide range of applications, including, for example, landmine and unexploded ordinance (UXO) detection using ground penetrating radar (GPR) \cite{CHND01}, hidden object imaging via foliage penetrating (FOPEN) radar \cite{XN01}, as well as human detection \cite{YLML06} and non-contact human vital sign monitoring using UWB radar \cite{SLL20}. Due to the large bandwidth, which can be over 10 GHz for an impulse UWB radar system, an analog-to-digital converter (ADC) with a high-sampling rate of over 20 GHz is needed at its receiver. However, a radar system using such an ADC, especially with high-resolution quantization, may be too expensive to be commercially viable. Indeed, high rate ADC with large quantization depth, even if available, can significantly increase the cost and power consumption of the UWB radar system. In contrast, an ADC with low-resolution quantization can be attractive due to its low-cost and low power consumption advantage and its ability to achieve ultra-high sampling rates \cite{ZHLLW18,ZHB19}. For instance, the NVA6100 impulse radar system \cite{Xethru}, a low-cost and low power consumption, single-chip UWB radar from Novelda, utilizes the so-called Continuous Time Binary Value (CTBV) technology \cite{Xethru, HWLLM07} to achieve a very high sampling rate of 39 GHz and a 13-bit quantization resolution with a simple circuit design. CTBV is an efficient one-bit sampling strategy, which obtains its signed measurements via comparing the received signal to a known threshold varying with slow-time, i.e., varying from one pulse repetition interval (PRI) to another. High-precision samples can be obtained from these signed measurements via a simple digital integration (DI) method \cite{HWLLM07}. The affordable NVA6100 system can be used for diverse applications, including vital sign monitoring \cite{SLL20}, through-wall imaging and object tracking \cite{Xethru}. We refer to the CTBV-based UWB radar system in this paper as the one-bit UWB radar system.

One of the most significant challenges of ensuring the proper operations of UWB radar systems is to mitigate the severe radio frequency interferences (RFIs) they encounter since there are many competing users within the ultra-wideband frequency range they operate in. Typical RFI sources include FM radio transmitters, TV broadcast transmitters, cellular phones, and other radiation devices. Their operating frequency bands tend to overlap with those of the UWB radar systems \cite{KL95}. These RFI sources pose a significant hindrance to the proper operations of the UWB radar systems in terms of reduced signal-to-noise ratio (SNR) and degraded radar imaging quality. Therefore, effective RFI mitigation is critically important for the proper operations of the UWB radar systems.

RFI mitigation is a notoriously challenging problem since it is difficult to predict and model RFI signals accurately due to their dynamic range and diverse modulation schemes. Many RFI mitigation methods, such as RFI suppression via filtering techniques \cite{KL95,C91,SA93,VSPH10,LUASF01} and RFI extraction based on RFI estimation methods \cite{MPM97,ZWXB07,ZTBL13,RZLNS19}, have been developed for radar systems using high-precision ADCs. However, it appears that RFI mitigation for one-bit UWB radar systems has not been considered in the literature before and the existing high-resolution quantization based methods are not directly applicable.

In this paper, we introduce a RFI mitigation method for the one-bit UWB radar systems using the CTBV sampling technique, in particular the NVA6100 impulse radar system. Our main contributions can be summarized as follows:

1) We present a novel RFI mitigation framework for one-bit UWB radar systems, which obtain their measurements using a one-bit sampling strategy that varies the known quantization threshold with slow-time.

\begin{figure}[htbp]
\centering
\includegraphics[width=0.48\textwidth]{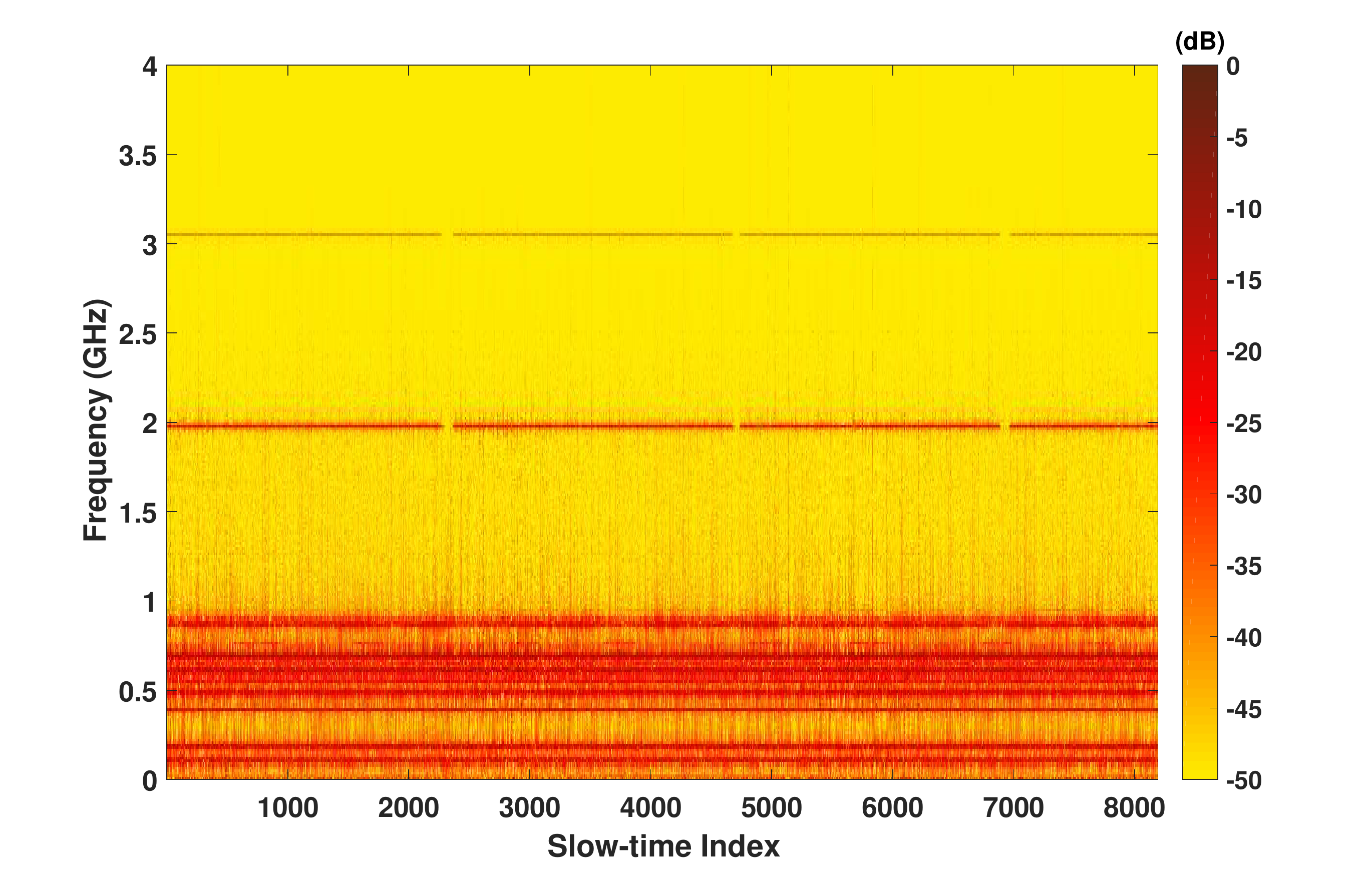}
\caption{An example of fast-time RFI spectrum vs. slow-time index, for the RFI-only data measured by the experimental ARL radar receiver.}
\label{RADAR2}
\end{figure}

2) We establish a proper data model for the RFI sources and extend the recently-developed majorization-minimization (MM) based 1bMMRELAX \cite{RZLS19} method for sinusoidal parameter estimation, i.e., for single-PRI based signed measurements, to deal with multiple-PRI based signed measurements. Specifically, since the desired UWB radar echo signals are relatively weak with a flat spectrum and the RFI sources are typically strong with sparse, narrow spectral peaks in the fast-time frequency domain (see Figure \ref{RADAR2}, for example), we consider modeling the RFI signal as a sum of sinusoids. The sinusoidal frequencies are assumed fixed within the coherent processing interval (CPI). The extended 1bMMRELAX algorithm \cite{ZRGL19} can be used to obtain the maximum likelihood (ML) estimates of the parameters of the RFI sources from the multiple-PRI based signed measurements.

3) To further reduce the computational cost of the extended 1bMMRELAX, we also devise an Alternating Direction Method of Multipliers (ADMM) \cite{BPCPE11} based fast frequency initialization method which exploits the sparsity of the RFI spectrum.

4) Since the number of RFI sources, i.e., the model order of RFI signals, is unknown, we extend the single-PRI based 1bBIC \cite{LZLS18} to the multiple-PRI based cases. We use the extended 1bMMRELAX with the extended 1bBIC to simultaneously estimate the RFI parameters and determine the number of RFI sources.

5) We model the desired UWB radar echoes as sparse impulses in the fast-time domain due to the sparsity of strong targets, and introduce an ADMM-based sparse method to efficiently and effectively recover the desired UWB radar echoes based on the estimated RFI parameters.

6) Both simulated and measured RFI examples are presented in this paper to demonstrate the effectiveness of the proposed methods, especially when the RFI problem is severe.

The rest of this paper is organized as follows. In Section \ref{PF}, we introduce the one-bit UWB radar system using the CTBV sampling technique and formulate the RFI mitigation problem for the system. Next, in Section \ref{1bMMRELAX}, we present the extended 1bMMREALX algorithm along with the fast frequency initialization method and the extended 1bBIC to estimate the RFI parameters and determine the number of RFI sources from multiple-PRI based signed measurements obtained by using the CBTV sampling strategy. Then, using the estimated RFI parameters, we introduce the ADMM based sparse echo signal recovery method in Section \ref{echo_recover}. Finally, in Section \ref{sec:exp}, we provide both simulated and experimental results to demonstrate the effectiveness of the proposed algorithm for RFI mitigation for one-bit UWB radar systems.

\emph{Notation:} We denote vectors and matrices by boldface lower-case and upper-case letters, respectively. $(\cdot)^T$ denotes the transpose operation. $\hat{(\cdot)}$ refers to the estimated result of the related value. ${\bf X}\in\mathbb{R}^{N\times M}$ denotes a real-valued $N\times M$ matrix and ${\bf x}\in\mathbb{R}^N$ denotes a real-valued vector with $N$ elements. ${\bf X}[n,m]$ means the $(n,m)$th element of matrix ${\bf X}$. ${\bf X}[n, :]$ and ${\bf X}[:,m]$ means the $n$th row and $m$th column of the matrix ${\bf X}$, respectively. ${\bf x}[n]$ denotes the $n$th element of the vector ${\bf x}$. For a matrix or a vector, $||\cdot||_p$ means the $\ell_p$ element-wise norm of this matrix or vector, i.e., $||{\bf X}||_p = (\sum_{m=1}^M\sum_{n=1}^N|{\bf X}[n,m]|^p)^{1/p}$ or $||{\bf x}||_p = (\sum_{n=1}^N|{\bf x}[n]|^p)^{1/p}$. $||{\bf X}||_{1,2} = \sum_{n=1}^N||{\bf X}[n,:]||_2$ denotes the $\ell_{1,2}$ norm of the matrix ${\bf X}$. ${\bf I}_N$ denotes the $N\times N$ identity matrix.
\section{Problem Formulation}\label{PF}
\subsection{One-bit UWB Radar System}\label{sec:system}
\begin{figure}[htbp]
\centering
\subfloat[]{\includegraphics[width=0.5\textwidth]{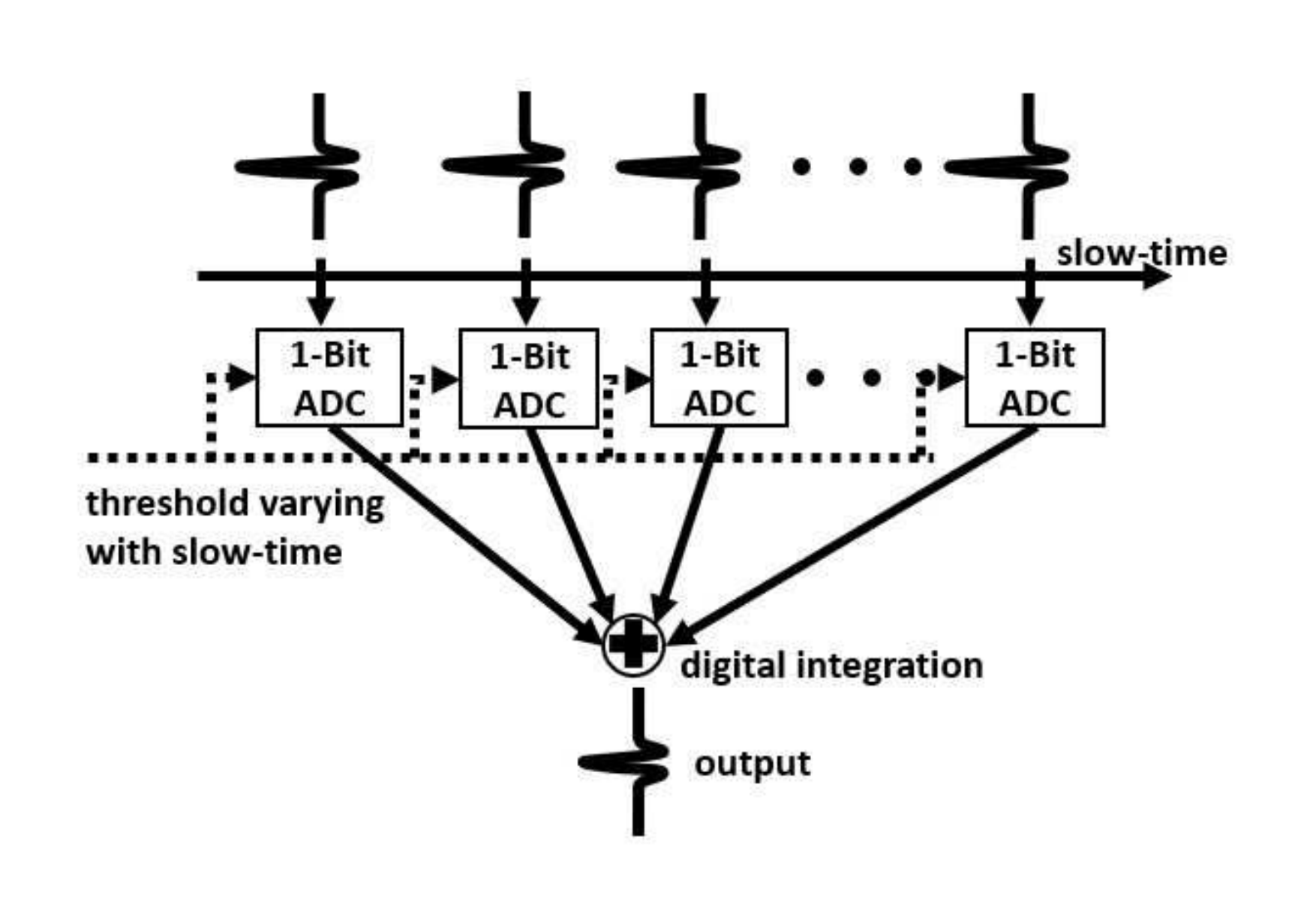}}\\
\subfloat[]{\includegraphics[width=0.5\textwidth]{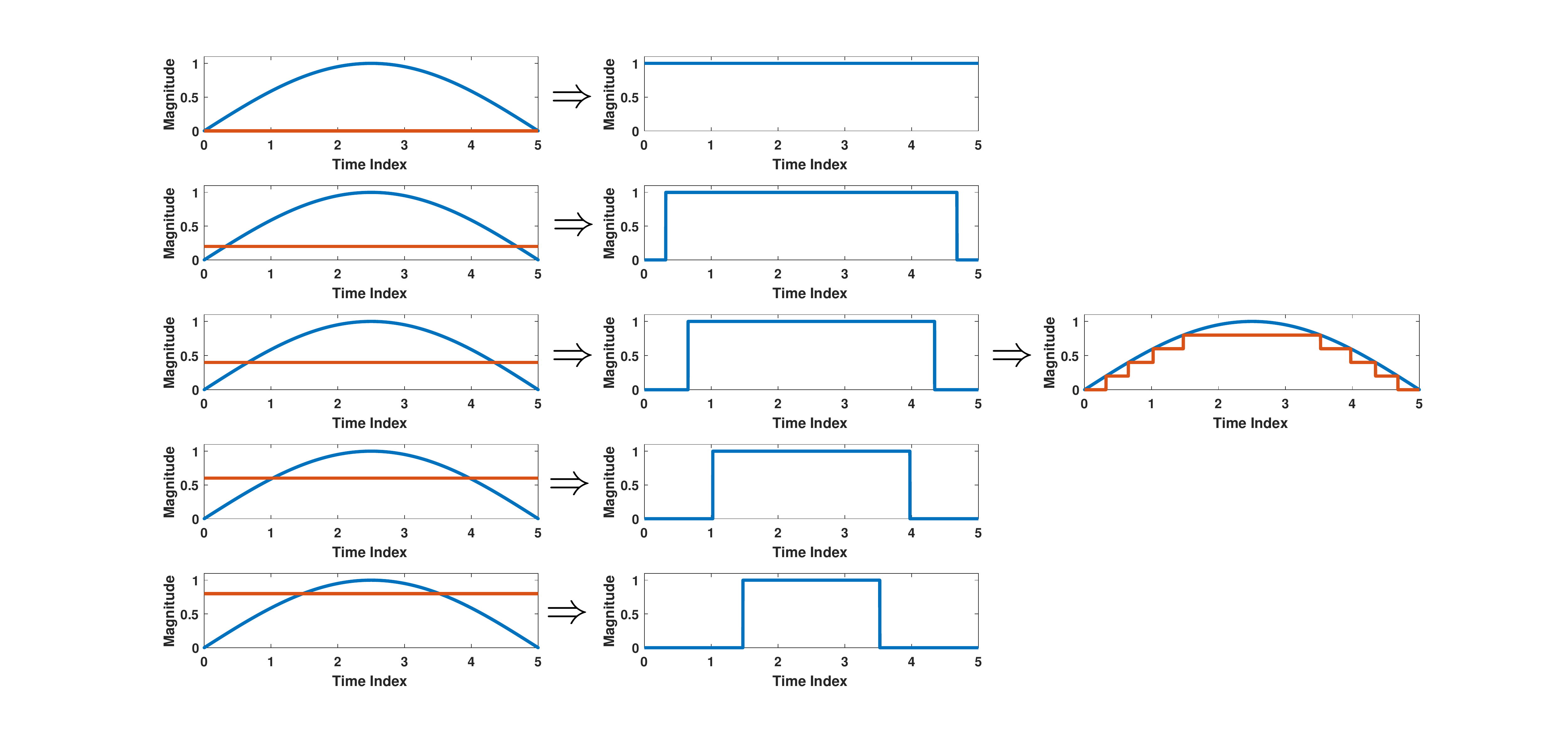}}
\caption{a) Structure of the receiver of a one-bit UWB radar system, b) an illustration of the DI method.}
\label{fig:structure}
\end{figure}

Consider an exemplary one-bit UWB radar system, the NVA6100 system, which is a recently-developed UWB radar system on a single chip with rather simple circuit designs \cite{Xethru}. It can achieve an ultra-high sampling frequency of 39 GHz and a 13-bit quantization resolution with low-cost and low power consumption hardware by using the CTBV sampling scheme, a kind of one-bit sampling technique. The radar transmits a super-narrow pulse repeatedly and uses a one-bit ADC with different thresholds to obtain the signed measurements. More specifically, each reflected signal will be compared with a known threshold, which varies uniformly over different PRIs within the CPI, and whether the sampling data is larger or smaller than the threshold will be recorded. Thus, in the absence of RFI and other interferences, the signed measurement matrix ${\bf Y}\in\mathbb{R}^{N\times M}$ obtained by NVA6100 can be expressed as follows:
\begin{equation}
{\bf Y} = {\rm sign}({\bf S}-{\bf H}),
\end{equation}
where $N$ and $M$ denote the number of fast-time samples per PRI and the number of PRIs or slow-time samples within the CPI, respectively, ${\bf S}$ denotes the desired radar echo signal and ${\bf H}$ denotes the known threshold matrix, whose columns vary linearly with slow-time, i.e., ${\bf H}[n,m] = -h+2(m-1)h/(M-1), h>0, n = 1,2,\cdots, N, m = 1,2,\cdots,M$. ${\rm sign}(\cdot)$ is the element-wise sign operator defined as:
\begin{equation}
{\rm sign}(x) = \begin{cases}1, & x\geq 0, \\ -1, & x<0.\end{cases}
\end{equation}
With the assumption that the reflected signal will be the same, i.e., ${\bf S}[:,1] = {\bf S}[:,2] = \cdots = {\bf S}[:,M]={\bf s}$, within a small time window, for example, within a CPI, the high-precision measurements can be obtained by using the simple digital integration (DI) method \cite{HWLLM07} from the signed measurement matrix ${\bf Y}$. The output of the one-bit system using the DI method, $\hat{\bf s}^{\rm DI}$, can be written in the following form:
\begin{equation}\begin{split}
\hat{\bf s}^{\rm DI}[n]=\left[\Delta h\sum_{m=1}^M\frac{1}{2}({\bf Y}[n,m]+1)\right]-h-\Delta h,\\
\Delta h = 2h/(M-1), n = 1,\dots,N.
\end{split}\end{equation}
The structure of the NVA6100 receiver and the procedure of the DI method are shown in Figure \ref{fig:structure}.
The DI method assumes no interference or weak interferences, and thus it cannot provide satisfactory performance for strong RFI mitigation. Since strong RFI problems exist in practical applications, an effective RFI mitigation technique is needed for the proper operations of the one-bit UWB radar systems.
\subsection{Data Model}
In the presence of RFI and other noise and disturbances, the signed measurement matrix ${\bf Y}$ can be written as follows:
\begin{equation}\label{signal}
\begin{split}
&{\bf Y} = {\rm sign}({\bf R}_{\bm\theta}+{\bf S}+{\bf E}-{\bf H}),\\
&{\bf S}[:,1] = {\bf S}[:,2] = \cdots = {\bf S}[:,M]={\bf s},
\end{split}
\end{equation}
where ${\bf E}$ denotes the noise and other disturbances and ${\bf R}_{\bm\theta}$ denotes the RFI matrix. By using the fact that the RFI sources tend to have strong, narrow peaks in the fast-time frequency domain and the frequencies of the RFI sources change only very slightly over the CPI (see Figure \ref{RADAR2}), the RFI sources can be modeled as a sum of sinusoids with their frequencies fixed over the slow-time within the CPI \cite{RZLNS19}. Thus, each element of ${\bf R}_{\bm\theta}$ can be expressed as follows:
\begin{equation}\begin{split}
{\bf R}_{\bm\theta}[n,m] &= \sum_{k=1}^K A_{k,m}\sin(\omega_k(n-1)+\phi_{k,m})\\
    &= \sum_{k=1}^K a_{k,m}\cos(\omega_k(n-1))+b_{k,m}\sin(\omega_k(n-1))\\
    & n = 1,\cdots, N, m = 1,\cdots,M,
\end{split}\end{equation}
where $K$ denotes the number of sinusoids or RFI sources, $\omega_k\in[0,\pi)$ denotes the frequency of the $k$th RFI source, and $A_{k,m}\in\mathbb{R}^{+}$ and $\phi_{k,m}
\in[0,2\pi)$ denote the amplitude and phase of the $k$th RFI source during the $m$th PRI, respectively. The unknown parameter vector of the RFI is denoted by ${\bm\theta} = [a_{1,1},b_{1,1},\dots,a_{1,M},b_{1,M},
\omega_1,\dots,a_{K,1},b_{K,1},\dots,a_{K,M}, \\ b_{K,M},\omega_K]^T \in\mathbb{R}^{(2M+1)K}$ with $a_{k,m}=A_{k,m}\sin\phi_{k,m}\in\mathbb{R}$ and $b_{k,m}=A_{k,m}\cos\phi_{k,m} \in\mathbb{R}$. Our goal is to recover the desired radar echo vector ${\bf s}$ from the signed measurement matrix ${\bf Y}$ while mitigating the impact of the RFI.
\section{1bMMRELAX For RFI Parameter Estimation}\label{1bMMRELAX}
\subsection{Maximum likelihood estimation}
We first assume that the desired UWB radar echoes together with the noise and other disturbances, i.e., ${\bf S}+{\bf E}$, obey i.i.d. Gaussian distribution with zero-mean and unknown variance $\sigma^2$. The numerical and experimental examples in Section \ref{sec:exp} show that the proposed algorithm is robust to this assumption. We consider the maximum likelihood (ML) estimator for the RFI parameter estimation problem due to its desirable properties including consistency and asymptotic efficiency. The ML estimate of the parameter vector ${\bm\beta}=[{\bm\theta}^T,\sigma]^T$ can be obtained by minimizing the following negative log-likelihood function \cite{GXLS16}:
\begin{equation}\label{neg-log-like}\begin{split}
& \hat{\tilde{\bm\beta}}= \arg\min_{\tilde{\bm\beta}} l({\tilde{\bm\beta}})\\
 &=\arg\min_{\tilde{\bm\beta}}\!\sum_{m=1}^M\sum_{n=1}^{N}-\log\!\Bigg[\!\Phi\!\Bigg(\!\!{\bf Y}[n,m]\!\Bigg(\!\!\sum_{k=1}^K\! \tilde{a}_{k,m}\cos(\omega_k(n\!-\!1))\\
 &+\!\tilde{b}_{k,m}\sin(\omega_k(n\!-\!1))\!\!-\!\!\lambda{\bf H}[n,m]\!\Bigg)\!\Bigg)\!\Bigg],
\end{split}\end{equation}
where $\Phi(x)$ denotes the cumulative distribution function of
the standard normal distribution, $\lambda=\frac{1}{\sigma}$, $\tilde{a}_{k,m}=\frac{a_{k,m}}{\sigma}$, $\tilde{b}_{k,m}=\frac{b_{k,m}}{\sigma}$, and $\tilde{\bm\beta}=[\tilde{\bm\theta}^T,\lambda]^T$ is the modified unknown parameter vector with $\tilde{\bm\theta}=[\tilde{a}_{1,1}, \tilde{b}_{1,1},\dots,\tilde{a}_{1,M},\tilde{b}_{1,M},\omega_1,\dots, \tilde{a}_{K,1},\tilde{b}_{K,1},\dots,\tilde{a}_{K,M},\\ \tilde{b}_{K,M},\omega_K]^T$. Note that the cost function in (\ref{neg-log-like}) is highly nonlinear and non-convex and is difficult to optimize with respect to ${\bm\beta}$.

Let ${\bm\omega} = [\omega_1,\omega_2,\dots,\omega_K]^T$ be a vector composed of the frequencies of ${\bf R}_{\bm\theta}$. For given ${\bm\omega}$, the above optimization problem is convex with respect to $\{\tilde{a}_{k,m}\}_{k=1,M=1}^{K,M}$, $\{\tilde{b}_{k,m}\}_{k=1,M=1}^{K,M}$ and $\lambda$ (see Section 3.54 of \cite{BV09} for a detailed proof). Thus, the ML problem can be solved by first performing a $K$-dimensional search over ${\bm\omega}$ on the feasible space of frequencies $[0,\pi)^K$ and then using globally optimal algorithms, for example, the Newton's Method \cite{BV09}, to compute the corresponding optimal values $\{\hat{\tilde{a}}_{k,m}\}_{k=1,M=1}^{K,M}$, $\{\hat{\tilde{b}}_{k,m}\}_{k=1,M=1}^{K,M}$ and $\hat{\lambda}$ as functions of ${\bm\omega}$. The details on the implementation of the $K$-dimensional frequency search can be found in \cite{GXLS17}.

Note that as the number of sinusoids $K$ and the size of the signed measurement matrix increase, this method can become computationally prohibitive and thus, more efficient algorithms, such as the relaxation based methods in \cite{RZLS19,GXLS17}, may be considered.
\subsection{1bREALX and 1bMMRELAX}\label{sec:1bRELAX_1bMMRELAX}
Inspired by the RELAX algorithm \cite{LS96}, which is a conceptually and computationally simple method for sinusoidal parameter estimation for infinite-precision data, two algorithms, 1bRELAX \cite{GXLS17} and 1bMMRELAX \cite{RZLS19}, were recently proposed as relaxation-based approaches to estimate the sinusoidal parameters from single-PRI based signed measurements by maximizing the likelihood function iteratively. Since 1bRELAX and 1bMMRELAX are both cyclic algorithms, their local convergence is guaranteed under wild conditions \cite{Z69}.

We extend below 1bRELAX and 1bMMRELAX to the case of multiple-PRI based signed measurements. We first extend the 1bRELAX algorithm to the multiple-PRI cases and the detailed steps are depicted in Table \ref{alg:1bRELAX}. 1bRELAX maximizes the likelihood function iteratively by conducting a sequence of one-dimensional frequency searches on $[0,\pi)$ rather than the $K$-dimensional search on $[0,\pi)^K$ used by the ML method, and therefore greatly reduces the computational cost. However, since the frequency update procedure is still implemented by means of an exhaustive search and a large number of iterations is required, 1bRELAX is still rather time-consuming.

\begin{table}[!t]
\caption{1bRELAX for Multiple-PRI Based Case}\label{alg:1bRELAX}
\centering
\resizebox{0.48\textwidth}{!}{
\begin{tabular}{cl}
\hline
1:&\textbf{Input:} Signed measurement matrix ${\bf Y}$, the desired or estimated \\
&model order $\hat{K}$, and the maximum number of update iteration $T_R$.\\
2:&Assume $\tilde{K}=1$. Obtain $\{\{\hat{\tilde{a}}_{1,m},\hat{\tilde{b}}_{1,m}\}_{m=1}^M,\hat{\omega}_1\}$ and $\hat{\lambda}$ by\\
&solving $(\ref{neg-log-like})$ via the exhaustive search (over $\omega_1$).\\
3:&\textbf{for} $\tilde{K}=2:\hat{K}$\\
4:&\quad$t=0$\\
5:&\quad Repeat:\\
6:&\quad\quad Obtain $\{\{\hat{\tilde{a}}_{\tilde{K},m},\hat{\tilde{b}}_{\tilde{K},m}\}_{m=1}^M,\hat{\omega}_{\tilde{K}}\}$ by solving (\ref{neg-log-like})\\
&\quad\quad via the exhaustive search with $\{\{\tilde{a}_{q,m},\tilde{b}_{q,m}\}_{m=1}^M,\omega_q\}_{q=1}^{\tilde{K}-1}$ \\
&\quad\quad and $\lambda$ replaced by their most recent estimates\\
&\quad\quad $\{\{\hat{\tilde{a}}_{q,m},\hat{\tilde{b}}_{q,m}\}_{m=1}^M,\hat{\omega}_q\}_{q=1}^{\tilde{K}-1}$ and $\hat{\lambda}$;\\
7:&\quad\quad Redetermine $\{\{\hat{\tilde{a}}_{1,m},\hat{\tilde{b}}_{1,m}\}_{m=1}^M,\hat{\omega}_1\}$ and $\hat{\lambda}$ by solving (\ref{neg-log-like}) \\
&\quad\quad via the exhaustive search with $\{\{\tilde{a}_{q,m},\tilde{b}_{q,m}\}_{m=1}^M,\omega_q\}_{q=2}^{\tilde{K}}$ \\
&\quad\quad replaced by their most recent estimates \\ &\quad\quad $\{\{\hat{\tilde{a}}_{q,m},\hat{\tilde{b}}_{q,m}\}_{m=1}^M,\hat{\omega}_q\}_{q=2}^{\tilde{K}}$;\\
8:&\quad\quad\textbf{if} $\tilde{K}>2$\\
9:&\quad\quad\quad\textbf{for} $k=2:\tilde{K}-1$\\
10:&\quad\quad\quad\quad Update $\{\{\tilde{a}_{k,m},\tilde{k}_{q,m}\}_{m=1}^M,\omega_k\}$ by by solving (\ref{neg-log-like}) via\\
&\quad\quad\quad\quad the exhaustive search with $\{\{\tilde{a}_{q,m},\tilde{b}_{q,m}\}_{m=1}^M,\omega_q\}_{q=1,q\neq k}^{\tilde{K}}$ \\
&\quad\quad\quad\quad and $\lambda$ replace by their most recent estimates\\
&\quad\quad\quad\quad $\{\{\hat{\tilde{a}}_{q,m},\hat{\tilde{b}}_{q,m}\}_{m=1}^M,\hat{\omega}_q\}_{q=1,q\neq k}^{\tilde{K}}$ and $\hat{\lambda}$.\\
11:&\quad\quad\quad\textbf{end}\\
12:&\quad\quad\textbf{end}\\
13:&\quad\quad $t=t+1$;\\
14:&\quad Until practical convergence or $t$ reaches the maximum number $T_R$.\\
15:&\textbf{end}\\
16:&\textbf{Output:} $\{\hat{a}_{k,m}\}_{k=1,m=1}^{\hat{K},M}=\{\hat{\tilde{a}}_{k,m}\}_{k=1,m=1}^{\hat{K},M}/{\hat\lambda}$,\\
&$\{\hat{b}_{k,m}\}_{k=1,m=1}^{\hat{K},M}=\{\hat{\tilde{b}}_{k,m}\}_{k=1,m=1}^{\hat{K},M}/{\hat\lambda}$, $\{\omega_k\}_{k=1}^{\hat{K}}$,and $\hat{\lambda}$.\\
\hline
\end{tabular}}
\end{table}
To enhance the computational efficiency of 1bRELAX, the majorization-minimization \cite{HL04} based 1bMMRELAX is introduced in \cite{RZLS19} for the single-PRI based cases. By using the MM technique, 1bMMRELAX transforms the likelihood maximization problem in each step into a sequence of simple infinite-precision sinusoidal parameter estimation problems, which can be efficiently solved via simple FFT operations. We next extend the single-PRI based 1bMMRELAX algorithm to the multiple-PRI based case, and we still refer to the extended algorithm as 1bMMRELAX for simplicity \cite{ZRGL19}.

We start by applying the MM-based method to minimize the negative log-likelihood function $l(\tilde{\bm\beta})$ in (\ref{neg-log-like}). The majorizing function can be obtained by using Lemma 1 in \cite{RZLS19}. The optimization problem at the $(i+1)$th MM iteration that minimizes the majorizing function at $\tilde{\bm\beta}^i=[(\tilde{\bm\theta}^i)^T,\lambda^i]^T$, which is the estimate of $\tilde{\bm\beta}$ obtained at the $i$th MM iteration, can be simplified as:
\begin{equation}\begin{split}\label{MM}
&\min_{\tilde{\bm\theta},\lambda} G\!\left(\!\tilde{\bm\theta},\lambda|\tilde{\bm\beta}^i\!\right)
\!\!=\!\!\!\!\sum_{m=1}^M\!\sum_{n=1}^{N}\!\!
\left[{\bf R}_{\tilde{\bm\theta}}[n,m]\!-\!\lambda{\bf H}[n,m]\!-\!\tilde{\bf Z}_{\tilde{\bm\beta}^i}[n,m]\right]^2,\\
\end{split}\end{equation}
where $\tilde{\bf Z}_{\tilde{\bm\beta}^i}[n,m]={\bf Y}[n,m]\!\left({\bf X}_{\tilde{\bm\beta}^i}[n,m]\!-\!f'({\bf X}_{\tilde{\bm\beta}^i}[n,m])\right)$ is the $(n,m)$th element of an auxiliary matrix $\tilde{\bf Z}_{\tilde{\bm\beta}^i}$, with $f(x)\!=\!-\log\Phi(x)$, and ${\bf X}_{\tilde{\bm\beta}^i}[n,m]={\bf Y}[n,m]\left({\bf R}_{\tilde{\bm\theta}^i}[n,m]\!-\!\lambda^i{\bf H}[n,m]\right)$, $n=1,\cdots,N$, $m=1,\cdots,M$.

The minimization problem in (\ref{MM}) can be conveniently solved by using a cyclic algorithm \cite{SS04}. In the cyclic algorithm, we conduct the following two steps iteratively: (1) minimize $G\left(\tilde{\bm\theta},\lambda|\tilde{\bm\beta}^i\right)$ with respect to $\lambda$ for fixed $\tilde{\bm\theta}$ and (2) minimize $G\left(\tilde{\bm\theta},\lambda|\tilde{\bm\beta}^i\right)$ with respect to $\tilde{\bm\theta}$ for given $\lambda$. The closed-form solution of the first step can be readily obtained as follows:
\begin{equation}\begin{split}
\lambda^{i+1}_{j+1}
\!\!=\!\!\max\!\!\left(\!\!0,\!\frac{\sum_{m=1}^M\!\!{\bf H}^T[:,m]\!\!\left[\!{\bf R}_{\tilde{\bm\theta}^{i+1}_j}[:,m]\!\!-\!\!\tilde{\bf Z}_{\tilde{\bm\beta}^i}[:,m]\!\right]}{||{\bf H}||_2^2}\!\right),
\end{split}\end{equation}
where the subscript $j$ denotes the iteration number in the cyclic minimization performed at the $i$th MM iteration. In the second step of the cyclic algorithm, by regarding $\{{\bf V}[n,m] = \lambda^{i+1}_{j+1}{\bf H}[n,m]+\tilde{\bf Z}_{\tilde{\bm\beta}^i}[n,m]\}_{n=1,m=1}^{N,M}$ as the input data, the minimization problem with respect to $\tilde{\bm\theta}$ can be interpreted as the infinite-precision direction-of-arrival (DOA) estimation problem encountered in array processing \cite{LZS97}:
\begin{equation}\label{equ:I-RELAX}
\begin{split}
&\min_{\omega} -\sum_{m=1}^M|{\bf a}(\omega){\bf V}[:, m]|^2,\\
&{\bf a}(\omega)=[1, e^{-j\omega},\dots,e^{-j\omega(N-1)}].
\end{split}
\end{equation}
Then $G\left(\tilde{\bm\theta},\lambda^{i+1}_{j+1}|\tilde{\bm\beta}^i\right)$ can be decreased efficiently by using the infinite-precision RELAX algorithm in \cite{LZS97} for the multiple-PRI based signals. The infinite-precision RELAX in \cite{LZS97} can be easily implemented by first using an $N_1$-point ($N_1>N$) zero-padded FFT for each PRI and followed by the subsequent fine search of minimizing (\ref{equ:I-RELAX}) via the Matlab fminbnd function over the interval $[\hat{\omega}_k^{\rm FFT}-\frac{\pi}{N_1},\hat{\omega}_k^{\rm FFT}+\frac{\pi}{N_1}]$, where $\hat{\omega}_k^{\rm FFT}$ is the frequency estimate of the $k$th RFI source obtained by using FFT. Since there exists a simple closed-form solution for $\lambda$, we re-determine $\lambda$ after updating each sinusoid.

The MM approach for solving the optimization problem in (\ref{neg-log-like}) is summarized in Table \ref{alg:MM}. Then, the 1bMMRELAX algorithm is obtained by replacing the update procedure of the extended 1bRELAX (see Steps $7-14$ of Table \ref{alg:1bRELAX}) by the MM steps proposed in Table \ref{alg:MM}. When initializing the MM approach via the exhaustive search, however, a $2M$-dimensional convex optimization problem needs to be solved $N$ times (see Step 6 of Table \ref{alg:1bRELAX}). For large $M$, this step is still computationally expensive and therefore a faster frequency initialization algorithm is needed.
\begin{table}[!t]
\caption{MM Steps for Solving (\ref{neg-log-like})}\label{alg:MM}
\centering
\resizebox{0.48\textwidth}{!}{
\begin{tabular}{cl}
\hline
1:&\textbf{Input:} Signed measurement matrix ${\bf Y}$, known threshold matrix ${\bf H}$,\\
& initialization $\tilde{\bm\theta}^0$ and $\lambda^0$, maximum number of MM iterations $T_M$,\\
& maximum number of inner loop iterations $T_C$, model order $K$,\\
& and $i=0$.\\
2:&Repeat:\\
3:&\quad Update: $\tilde{\bm\beta}^i=[\tilde{\bm\theta}^T,\lambda^i]^T$,\\
4:&\quad $\tilde{\bf Z}_{\tilde{\bm\beta}^i}[n,m]={\bf Y}[n,m]\!\left({\bf X}_{\tilde{\bm\beta}^i}[n,m]\!-\!f'({\bf X}_{\tilde{\bm\beta}^i}[n,m])\right)$\\
&\quad $n=1,\dots,N, m = 1,\dots,M$,\\
5:&\quad $\tilde{\bm\theta}^{i+1}_0=\tilde{\bm\theta}^{i+1},j=0,k=K$.\\
6:&\quad Repeat:\\
7:&\quad\quad Update $\lambda^{i+1}_{j+1}
\!\!=\!\!\max\!\!\left(\!\!0,\!\frac{\sum_{m=1}^M\!\!\sum_{n=1}^{N}\!\!{\bf H}[n,m]\!\left[\!{\bf R}_{\tilde{\bm\theta}^{i+1}_j}[n,m]-\tilde{\bf Z}_{\tilde{\bm\beta}^i}[n,m]\!\right]}{\sum_{m=1}^M\sum_{n=1}^{N} {\bf H}^2[n,m]}\!\right);$\\
8:&\quad\quad ${\bf V}^k[n,m]=\tilde{\bf Z}_{\tilde{\bm\beta}^i}[n,m]+\lambda^{i+1}_{j+1}{\bf H}[n,m]-\sum_{q=1,q\neq k}^K$\\
&\quad\quad
$\{\tilde{a}_{q,m,j}^{i+1}\cos(\omega_{q,j}^{i+1}(n-1))+\tilde{b}_{q,m,j}^{i+1}\sin(\omega_{q,j}^{i+1}(n-1))\}$\\
&\quad\quad $n=1,\dots,N, m = 1,\dots,M$;\\
9:&\quad\quad Update $\{\{\tilde{a}_{k,m,j+1}^{i+1},\tilde{b}_{k,m,j+1}^{i+1}\}_{m=1}^M,\omega_{k,j+1}^{i+1}\}$ by using infinite-\\
&\quad\quad precision RELAX \cite{LZS97} for multiple-PRI based signals on \\
&\quad\quad $\{{\bf V}^k[n,m]\}_{n=1,m=1}^{N,M}$.\\
10:&\quad\quad $\tilde{\bm\theta}_{j+1}^{i+1}=[\tilde{a}_{1,1,j+1}^{i+1}, \tilde{b}_{1,1,j+1}^{i+1},\dots,\tilde{a}_{1,M,j+1}^{i+1},\tilde{b}_{1,M,j+1}^{i+1},\omega_{1,j+1}^{i+1},$\\
&\quad\quad $\dots, \tilde{a}_{K,1,j+1}^{i+1},\tilde{b}_{K,1,j+1}^{i+1},\dots,\tilde{a}_{K,M,j+1}^{i+1},\tilde{b}_{K,M,j+1}^{i+1},\omega_{K,j+1}^{i+1}]^T$.\\
11:&\quad\quad $k={\rm mod}(k,K)+1$, where ${\rm mod}(\cdot)$ denotes the modulo operation;\\
12:&\quad\quad $j=j+1$;\\
13:&\quad Until practical convergence or $j$ reaches the maximum number $T_C$.\\
14:&\quad $i=i+1$.\\
15:&Until practical convergence or $i$ reaches the maximum number $T_M$.\\
16:&\textbf{Output:} $\hat{\tilde{\bm\theta}}$ and $\hat{\tilde{\lambda}}$.\\
\hline
\end{tabular}}
\end{table}
\subsection{Fast frequency initialization}
To further improve the computational efficiency of the extended 1bMMRELAX, we devise a fast frequency initialization algorithm to estimate the frequencies of the RFI sources by exploiting the sparse property of the RFI spectrum.

Let $\{w_q = \frac{(q-1)\pi}{Q}\}_{q=1}^{Q}$ denote a grid that covers $[0,\pi)$. Assuming that the grid is fine enough such that the frequencies (normalized by the sampling frequency) corresponding to the RFI sources are on this grid (or practically, close to the grid) \cite{RZLNS19}, the $(n,m)$th element of the RFI signal ${\bf R} $ can then be rewritten as follows:
\begin{equation}
{\bf R}[n,m] = \sum_{q=1}^{Q}\!\breve{a}_{q,m}\cos(w_q(n-1))+\breve{b}_{q,m}\sin(w_q(n-1)).
\end{equation}
Then, denote
\begin{equation}\begin{split}\label{F_trans}
&{\bf F}=\\
&\!\!\begin{bmatrix}
1&\dots&1&0&\dots&0\\
\cos(w_1)&\dots&\cos(w_{\!Q})&\sin(w_1)&\dots&\sin(w_{\!Q})\\
\vdots&\vdots&\vdots&\vdots&\vdots&\vdots\\
\!\cos(\!w_{\!1}\!(\!N\!\!-\!\!1\!)\!)\!\!\!\!\!\!&\!\!\!\!\!\!\dots\!\!\!\!\!\!&\!\!\!\!\!\!\cos(\!w_{\!Q}\!(\!N\!\!-\!\!1\!)\!)&\sin(\!w_{\!1}\!(\!N\!-\!\!1\!)\!)
\!\!\!\!\!\!&\!\!\!\!\!\!\dots\!\!\!\!\!\!&\!\!\!\!\!\!\sin(\!w_{\!Q}\!(\!N\!\!-\!\!1\!)\!)\!\!\\
\end{bmatrix}\!\!,\\
\end{split}\end{equation}
and
\begin{equation}\begin{split}
{\bf A}=\begin{bmatrix}
\breve{a}_{1,1}&\cdots&\breve{a}_{1,M}\\
\vdots&\ddots&\vdots\\
\breve{a}_{Q,1}&\cdots&\breve{a}_{Q,M}\\
\breve{b}_{1,1}&\cdots&\breve{b}_{1,M}\\
\vdots&\ddots&\vdots\\
\breve{b}_{Q,1}&\cdots&\breve{b}_{Q,M}\\
\end{bmatrix}.\\
\end{split}\end{equation}
With the assumption that ${\bf S}+{\bf E}$ obeys i.i.d. Gaussian distribution with zero-mean and unknown variance $\sigma^2$, we can establish an optimization problem as follows, by making use of the group sparsity of the RFI spectrum \cite{RZLNS19} and the negative log-likelihood function:

\begin{equation}\label{opti1}
\min_{\tilde{\bf A},\lambda}\ \zeta_1||\tilde{\bf A}||_{1,2}+||f({\bf Y}\odot({\bf F}\tilde{\bf A}-\lambda{\bf H}))||_1,
\end{equation}
where $\lambda = 1/\sigma$, $\tilde{\bf A} = {\bf A}/\sigma$, $\odot$ denotes the element-wise matrix product, $\zeta_1$ is a user-parameter controlling the balance between the two penalty terms, and $f(x)\!=\!-\log\Phi(x)$.

Since the optimization problem in (\ref{opti1}) is difficult to solve directly, similar to Section \ref{sec:1bRELAX_1bMMRELAX}, the MM technique is used to obtain a simplified solution. By using the MM technique, the update formula at the $(i+1)$th MM iteration can be written as:
\begin{equation}\begin{split}\label{opti1-MM}
\min_{\tilde{\bf A},\lambda}\ \zeta_1||\tilde{\bf A}||_{1,2}+||{\bf F}\tilde{\bf A} - \lambda{\bf H} - \tilde{\bf Z}_{\rm FI}^{i}||_2^2, \\
\tilde{\bf Z}_{\rm FI}^i = {\bf Y}\odot(\tilde{\bf X}_{\rm FI}^i - f'(\tilde{\bf X}_{\rm FI}^i)),\\
\tilde{\bf X}_{\rm FI}^i = {\bf Y}\odot({\bf F}\tilde{\bf A}^i-\lambda^i{\bf H}).
\end{split}\end{equation}

Note that in (\ref{opti1-MM}), we minimize the majorization function of (\ref{opti1}) at $\{\tilde{\bf A}^{i}, \lambda^{i}\}$, where $\{\tilde{\bf A}^{i}, \lambda^{i}\}$ is the estimate of $\{\tilde{\bf A}, \lambda\}$ obtained at the $i$th MM iteration. To solve (\ref{opti1-MM}) by ADMM \cite{BPCPE11} efficiently and effectively, we rewrite (\ref{opti1-MM}) as follows:
\begin{equation}\begin{split}\label{opti1-MM-ADMM}
&\min_{\tilde{\bf A},\tilde{\bf B}_{\rm FI},\lambda}\ \zeta_1||\tilde{\bf A}||_{1,2}+||{\bf F}\tilde{\bf B}_{\rm FI} - \lambda{\bf H} - \tilde{\bf Z}_{\rm FI}^{i}||_2^2, \\
&{\rm s.t.}\ \tilde{\bf A} = \tilde{\bf B}_{\rm FI}.
\end{split}\end{equation}
Then the augmented Lagrangian for (\ref{opti1-MM-ADMM}), with the Lagrange multiplier ${\bf \Upsilon}_{\rm FI}$, is given as:
\begin{equation}\begin{split}
&\mathcal{L}_{\rm FI}(\tilde{\bf A},\tilde{\bf B}_{\rm FI},\lambda, {\bf \Upsilon}_{\rm FI}) = \zeta_1||\tilde{\bf A}||_{1,2}+||{\bf F}\tilde{\bf B}_{\rm FI} - \lambda{\bf H} - \tilde{\bf Z}_{\rm FI}^{i}||_2^2 \\
&+ \sum_{m=1}^M {\bf \Upsilon}_{\rm FI}[:,m]^T(\tilde{\bf A}[:,m]-\tilde{\bf B}_{\rm FI}[:,m])+\frac{\rho_{\rm FI}}{2}||\tilde{\bf A}-\tilde{\bf B}_{\rm FI}||_2^2,
\end{split}\end{equation}
where $\rho_{\rm FI}>0$ is the penalty parameter. At each iteration of ADMM, we minimize $\mathcal{L}_{\rm FI}$ with respect to $\{\tilde{\bf A},\tilde{\bf B}_{\rm FI},\lambda, {\bf \Upsilon}_{\rm FI}\}$ sequentially. The detailed update steps of ADMM at the $(i+1)$th MM iteration are described as follows.
\subsubsection{Update of $\tilde{\bf A}$}
The subproblem with respect to $\tilde{\bf A}$ is:
\begin{equation}\begin{split}
\min_{\tilde{\bf A}}\ &\zeta_1||\tilde{\bf A}||_{1,2}+\sum_{m=1}^M {\bf \Upsilon}^i_{\rm FI}[:,m]^T(\tilde{\bf A}[:,m]-\tilde{\bf B}^i_{\rm FI}[:,m])\\
&+\frac{\rho_{\rm FI}}{2}||\tilde{\bf A}-\tilde{\bf B}^i_{\rm FI}||_2^2.
\end{split}\end{equation}
For given $\{{\bf \Upsilon}^i_{\rm FI}, \tilde{\bf B}^i_{\rm FI}\}$, the solution to this subproblem is:
\begin{equation}\begin{split}\label{opti1-ADMM-A}
\tilde{\bf A}^{i+1} = \frac{1}{\rho}{\rm diag}({\bf c})(\rho_{\rm FI}{\tilde{\bf B}^i_{\rm FI}}-{\bf \Upsilon}^i_{\rm FI}),\\
{\rm c}[n] = \max\left(0, 1-\frac{\zeta_1}{||\rho_{\rm FI}{\tilde{\bf B}^i_{\rm FI}}[n,:]-{\bf \Upsilon}^i_{\rm FI}[n,:]||_2}\right),
\end{split}\end{equation}
where ${\rm diag}(\cdot)$ denotes the diagonal matrix whose diagonal elements are composed of the corresponding vector.
\subsubsection{Update of $\lambda$}
The subproblem with respect to $\lambda$ is:
\begin{equation}
\min_{\lambda}\ ||{\bf F}\tilde{\bf B}^i_{\rm FI} - \lambda{\bf H} - \tilde{\bf Z}_{\rm FI}^{i}||_2^2.
\end{equation}
The solution to this subproblem is:
\begin{equation}\label{opti1-ADMM-lmbd}
\lambda^{i+1} = \max\left(\!0,\frac{\sum_{m=1}^M{\bf H}[:,m]^T({\bf F}\tilde{\bf B}^i_{\rm FI}[:,m] - \tilde{\bf Z}_{\rm FI}^{i}[:,m])}{||{\bf H}||_2^2}\!\right).
\end{equation}
\subsubsection{Update of $\tilde{\bf B}_{\rm FI}$}
With the estimated $\{\tilde{\bf A}^{i+1},  \lambda^{i+1}\}$ and ${\bf \Upsilon}^i_{\rm FI}$, the subproblem with respect to $\tilde{\bf B}_{\rm FI}$ is:
\begin{equation}\begin{split}
\min_{\tilde{\bf B}_{\rm FI}}&\ ||{\bf F}\tilde{\bf B}_{\rm FI} - \lambda^{i+1}{\bf H} - \tilde{\bf Z}_{\rm FI}^{i}||_2^2\\
&+\!\!\sum_{m=1}^M {\bf \Upsilon}^i_{\rm FI}[:,m]^T(\tilde{\bf A}^{i+1}[:,m]-\tilde{\bf B}_{\rm FI}[:,m])\!\\
&+\!\frac{\rho_{\rm FI}}{2}||\tilde{\bf A}^{i+1}-\tilde{\bf B}_{\rm FI}||_2^2.
\end{split}\end{equation}
The solution to this subproblem is:
\begin{equation}
\label{opti1-ADMM-B}
\begin{split}
\tilde{\bf B}^{i+1}_{\rm FI} = &({\bf F}^T{\bf F}+\rho_{\rm FI}{\bf I}_{2N})^{-1}\\
&(\lambda^{i+1}{\bf F}^T{\bf H}+{\bf F}^T\tilde{\bf Z}_{\rm FI}^{i}+{\bf \Upsilon}^i_{\rm FI}+\rho_{\rm FI}\tilde{\bf A}^{i+1}).
\end{split}
\end{equation}
\subsubsection{Update of ${\bf \Upsilon}_{\rm FI}$}
The Lagrange multiplier ${\bf \Upsilon}_{\rm FI}$ can be updated in a manner of gradient descent:
\begin{equation}\label{opti1-ADMM-U}
{\bf \Upsilon}^{i+1}_{\rm FI}\leftarrow{\bf \Upsilon}^i_{\rm FI}+\rho_{\rm FI}(\tilde{\bf A}^{i+1}-\tilde{\bf B}^{i+1}_{\rm FI}).
\end{equation}
\subsubsection{Stopping Criteria}\label{sec:ADMMSC1}
We stop the ADMM iterations when the primal, dual residuals and the $\lambda$ residual are all sufficiently small. At the $(j+1)$th iteration of ADMM, the primal residual is:
\begin{equation}\label{update_pr_FI}
r_{\rm priFI}^{j+1} = ||\tilde{\bf A}^{j+1}-\tilde{\bf B}_{\rm FI}^{j+1}||_2,
\end{equation}
the dual residual is:
\begin{equation}
r_{\rm dualFI}^{j+1} = \rho_{\rm FI}||\tilde{\bf B}_{\rm FI}^{j+1}-\tilde{\bf B}_{\rm FI}^{j}||_2,
\end{equation}
and the $\lambda$ residual is:
\begin{equation}\label{update_lr_FI}
r_{\rm \lambda FI}^{j+1} = ||\lambda^{j+1}-\lambda^j||_1/\lambda^{j+1}.
\end{equation}
The stopping criteria are:
\begin{equation}\label{sc1}
r_{\rm priFI}^{j}\leq \varepsilon_{\rm priFI}^{j}, r_{\rm dualFI}^{j}\leq \varepsilon_{\rm dualFI}^{j}\ {\rm and}\ r_{\rm \lambda FI}^{j}\leq \varepsilon_{\rm \lambda FI}^j,
\end{equation}
where $\varepsilon_{\rm priFI}^{j}, \varepsilon_{\rm dualFI}^{j}$ and $\varepsilon_{\rm \lambda FI}^j$ are tolerances for primal, dual and $\lambda$ residuals for the $j$th ADMM iteration, respectively. In our problem, $\varepsilon_{\rm \lambda FI}^j$ can be set as a small constant and $\varepsilon_{\rm priFI}^{j}, \varepsilon_{\rm dualFI}^{j}$ can be set as follows:
\begin{equation}\label{update_tol_FI}
\begin{split}
&\varepsilon_{\rm priFI}^{j} = \sqrt{2NM}\varepsilon_{\rm absFI} + \varepsilon_{\rm relFI}\max (||\tilde{\bf A}^j||_2,||\tilde{\bf B}_{\rm FI}^j||_2),\\
&\varepsilon_{\rm dualFI}^{j} = \sqrt{2NM}\varepsilon_{\rm absFI} + \varepsilon_{\rm relFI}||{\bf \Upsilon}_{\rm FI}^j||_2,
\end{split}\end{equation}
where $\varepsilon_{\rm absFI}$ and $\varepsilon_{\rm relFI}$ are absolute relative tolerances, respectively, which are set depending on the scale of the problem.
\subsubsection{Update of $\rho_{\rm FI}$}\label{sec:ADMMRHO1}
To improve the convergence rate and  reduce the influence of the initial value of the penalty parameter $\rho_{\rm FI}$, we update the penalty parameter for each ADMM iteration using the following updating scheme \cite{BPCPE11}:
\begin{equation}\label{update_rho_FI}
\rho_{\rm FI}^{j+1} = \begin{cases}2\rho_{\rm FI}^{j},& r_{\rm priFI}^{j} > 10r_{\rm dualFI}^{j}, \cr 0.5\rho_{\rm FI}^{j},& r_{\rm dualFI}^{j} > 10r_{\rm priFI}^{j}, \cr \rho_{\rm FI}^{j},& {\rm otherwise},\end{cases}
\end{equation}
where $\rho_{\rm FI}^{j}$ denotes the penalty parameter in $j$th ADMM iteration.

The fast frequency initialization can be summarized in Table \ref{alg:FF}.
\begin{table}[!t]
\caption{Fast Frequency Initialization (FI)}\label{alg:FF}
\centering
\resizebox{0.48\textwidth}{!}{
\begin{tabular}{cl}
\hline
1:&\textbf{Input:} Signed measurement matrix ${\bf Y}$, known threshold matrix ${\bf H}$,\\
& initial values $\tilde{\bf A}^0, \lambda^0$, and $\varepsilon_{\rm absFI},\varepsilon_{\rm relFI}, \varepsilon_{\rm \lambda FI}$.\\
2:&MM Repeat:\\
3:&\quad Initialize $\rho_{\rm FI}$ for ADMM iteration.\\
4:&\quad Construct $\tilde{\bf Z}_{\rm FI}^i$ based on $\tilde{\bf A}^i, \lambda^i$ by (\ref{opti1-MM}).\\
5:&\quad ADMM Repeat:\\
6:&\quad\quad Update $\tilde{\bf A}$ by (\ref{opti1-ADMM-A}); \\
7:&\quad\quad Update $\lambda$ by (\ref{opti1-ADMM-lmbd}); \\
8:&\quad\quad Update $\tilde{\bf B}_{\rm FI}$ by (\ref{opti1-ADMM-B}); \\
9:&\quad\quad Update ${\bf \Upsilon}_{\rm FI}$ by (\ref{opti1-ADMM-U}); \\
10:&\quad\quad Calculate $r_{\rm priFI},r_{\rm dualFI},r_{\rm \lambda FI}, \varepsilon_{\rm priFI}, \varepsilon_{\rm dualFI}$ by (\ref{update_pr_FI})$\sim$(\ref{update_lr_FI}) and (\ref{update_tol_FI});\\
11:&\quad\quad Update $\rho_{\rm FI}$ by (\ref{update_rho_FI}).\\
12:&\quad Until stopping criteria of ADMM iterations $(\ref{sc1})$ are satisfied.\\
13&Until stopping criteria of MM iterations are satisfied.\\
14:&\textbf{Output:} $\hat{\tilde{\bf A}}_{\rm FI}$ and $\hat{\lambda}_{\rm FI}$, $\hat{\bf A}_{\rm FI} = \hat{\lambda}_{\rm FI}\hat{\tilde{\bf A}}_{\rm FI}$.\\
\hline
\end{tabular}}
\end{table}
From $\hat{\bf A}_{\rm FI}$, we obtain $\mathring{\bf A}= \sqrt{\hat{\bf A}_{\rm FI}[1:Q,:]^2+\hat{\bf A}_{\rm FI}[Q+1:2Q,:]^2}$, where the square and square root are both element-wise operations. Suppose that the maximum possible model order is $K_{\max}$. Since the frequencies of the RFI sources are known to be not close to zero, the $K_{\max}$ fast frequency initializations of the RFI sources, referred to as $\{\hat{\omega}_k^{\rm FI}\}_{k=1}^{K_{\max}}$, correspond to the $K_{\max}$ row indices of $\mathring{\bf A}[2:Q,:]$ (note that the row index $1$ corresponds to the zero frequency) with the $K_{\max}$ largest $\ell_1$ norms. The $K_{\max}$ initial frequency estimates are then sorted based on arranging the corresponding $\ell_1$ norms in descending order.

For the $\tilde{K}$th step of 1bMMRELAX, the coarse initial frequency estimate of the $\tilde{K}$th sinusoid is taken as $\hat{\omega}_{\tilde{K}}^{\rm FI}$, in lieu of the one obtained via the exhaustive search. Also, $\hat{\lambda}_{\rm FI}$ can be used as an initial value of $\lambda$ in the first step of 1bMMRELAX.
\subsection{1bBIC}
For the case that the number of RFI sources, i.e., the true model order $K$, is unknown, we consider extending the single-PRI based one-bit Bayesian information criterion (1bBIC) in \cite{LZLS18} to the multiple-PRI based cases so that the extended 1bBIC can be used with the extended 1bMMRELAX to determine the RFI model order. Suppose that $\hat{\bm\beta}_{\tilde{K}}$ is parameter vector estimated by the extended 1bMMRELAX algorithm for an assumed model order $\tilde{K}$, where $1\leq \tilde{K}\leq K_{\max}$. It is proven in Appendix A that the 1bBIC cost function for the multiple-PRI based signed measurements has the following form:
\begin{equation}\begin{split}
&{\rm 1bBIC}({\tilde K})= \\
&-2\sum_{m=1}^{M}\sum_{n=1}^{N}\log \left(\Phi\left({\bf Y}[n,m]\frac{\hat{\bf R}_{\hat{\bm\theta}_{\tilde K}}[n,m]-{\bf H}[n,m]}{\hat\sigma}\right)\right)\\
&+{\tilde K}(3+2M)\log N.
\end{split}\end{equation}
The estimate $\hat{K}$ of $K$ is selected as the integer that minimizes the above 1bBIC cost function with respect to the assumed number of sinusoids $\tilde K$.
\section{Radar Echo Signal Recovery}\label{echo_recover}
Due to the sparsity of strong targets in a scene of interest, the desired UWB radar echo vector ${\bf s}$ is in general quite sparse. Thus, using the estimated RFI sources, ${\bf s}$ can be recovered by exploiting its sparsity and minimizing the corresponding negative log-likelihood function. Since the desired signal is invariant over all PRIs within the CPI, (\ref{signal}) can be rewritten approximately as follows:
\begin{equation}\begin{split}
{\bf Y}[:,m] &\approx {\rm sign}({\bf D}{\bm\gamma}-{\bf U}[:,m]), m=1,\dots,M,\\
{\bf U} &= {\bf H}-\hat{\bf R}_{\hat{\bm\theta}}\in\mathbb{R}^{N\times M},
\end{split}\end{equation}
where $\hat{\bf R}_{\hat{\bm\theta}}$ is the RFI estimate, ${\bf D}\in\mathbb{R}^{N\times N}$ denotes the dictionary whose columns are time-shifted digitized versions of the transmitted impulse, and ${\bm\gamma}\in\mathbb{R}^{N}$ is a sparse vector containing the information of the magnitudes and positions of the radar echoes. The estimate of ${\bm\gamma}$ can be obtained by solving the following convex optimization problem \cite{BB08}:
\begin{equation}\label{opti2}
\min_{\tilde{\bm\gamma},\lambda}\ \zeta_2 M||\tilde{\bm\gamma}||_1+\sum_{m=1}^M||f({\bf Y}[:,m]\odot({\bf D}\tilde{\bm\gamma}-\lambda{\bf U}[:,m]))||_1,
\end{equation}
where $\zeta_2$ is a user-parameter and $\tilde{\bm\gamma} = {\bm\gamma}/\sigma, \lambda = 1/\sigma$, and $f(x)= -\log\Phi(x)$.

Similar to (\ref{opti1}), the convex objective function (\ref{opti2}) can be minimized efficiently by using the MM and ADMM techniques. The optimization problem at the $(i+1)$th MM iteration of (\ref{opti2}) can be written as:
\begin{equation}\begin{split}\label{opti2-MM}
\min_{\tilde{\bm\gamma},\lambda}\ \zeta_2 M||\tilde{\bm\gamma}||_1
+ \sum_{m=1}^M ||{\bf D}\tilde{\bm\gamma}-\lambda{\bf U}[:,m]-\tilde{\bf Z}_{\rm ER}^{i}[:,m]||_2^2
\end{split}\end{equation}
where
\begin{equation}\begin{split}
&\tilde{\bf Z}_{\rm ER}^{i} = {\bf Y}\odot(\tilde{\bf X}_{\rm ER}^{i}-f'(\tilde{\bf X}_{\rm ER}^{i}))\\
&\tilde{\bf X}_{\rm ER}^{i}[:,m] = {\bf Y}[:,m]\odot({\bf D}\tilde{\bm\gamma}^i-\lambda^i{\bf U}[:,m]), m = 1,2,\dots, M.
\end{split}\end{equation}
Here, $\{\tilde{\bm\gamma}^i, \lambda^i\}$ denotes the estimate of $\{\tilde{\bm\gamma}, \lambda\}$ at the $i$th MM iteration. Then, (\ref{opti2-MM}) can be solved effectively and efficiently by ADMM. To use ADMM technique, we rewrite (\ref{opti1-MM}) as follows:
\begin{equation}\begin{split}\label{opti2-MM-ADMM}
&\min_{\tilde{\bm\gamma},\tilde{\bf B}_{\rm ER},\lambda}\ \zeta_2 M||\tilde{\bm\gamma}||_1\\
& + \sum_{m=1}^M ||{\bf D}\tilde{\bf B}_{\rm ER}[:,m]-\lambda{\bf U}[:,m]-\tilde{\bf Z}_{\rm ER}^{i}[:,m]||_2^2\\
&{\rm s.t.}\ \tilde{\bm\gamma} = \tilde{\bf B}_{\rm ER}[:,m], \ m = 1,2,\dots, M.
\end{split}\end{equation}
Then the augmented Lagrangian for (\ref{opti2-MM-ADMM}) with the Lagrange multiplier ${\bm\Upsilon}_{\rm ER}$ is given as:
\begin{equation}\begin{split}
&\mathcal{L}_{\rm ER}(\tilde{\bm\gamma},\tilde{\bf B}_{\rm ER},\lambda, {\bf \Upsilon}_{\rm ER}) =\zeta_2 M||\tilde{\bm\gamma}||_1\\
& + \sum_{m=1}^M ||{\bf D}\tilde{\bf B}_{\rm ER}[:,m]-\lambda{\bf U}[:,m]-\tilde{\bf Z}_{\rm ER}^{i}[:,m]||_2^2 \\
&+ \sum_{m=1}^M {\bf \Upsilon}_{\rm ER}[:,m]^T(\tilde{\bm\gamma}-\tilde{\bf B}_{\rm ER}[:,m])\\
& +\frac{\rho_{\rm ER}}{2}\sum_{m=1}^M||\tilde{\bm\gamma}-\tilde{\bf B}_{\rm ER}[:,m]||_2^2,
\end{split}\end{equation}
where $\rho_{\rm ER}>0$ is the penalty parameter and in each iteration of ADMM, we minimize $\mathcal{L}_{\rm ER}$ by updating $\{\tilde{\bm\gamma},\tilde{\bf B}_{\rm ER},\lambda, {\bf \Upsilon}_{\rm ER}\}$ sequentially. The detailed update steps are described as follows.
\subsubsection{Update of $\tilde{\bm\gamma}$}
Given $\{{\bf \Upsilon}^i_{\rm ER}, \tilde{\bf B}^i_{\rm ER}\}$, the subproblem with respect to $\tilde{\bm\gamma}$ is:
\begin{equation}\begin{split}
&\min_{\tilde{\bm\gamma}}\ \zeta_2 M||\tilde{\bm\gamma}||_1+\sum_{m=1}^M {\bf \Upsilon}^i_{\rm ER}[:,m]^T(\tilde{\bm\gamma}-\tilde{\bf B}^i_{\rm ER}[:,m])\\
&+\frac{\rho_{\rm ER}}{2}\sum_{m=1}^M||\tilde{\bm\gamma}-\tilde{\bf B}^i_{\rm ER}[:,m]||_2^2.
\end{split}\end{equation}
The solution to this subproblem is:
\begin{equation}\begin{split}\label{opti2-ADMM-A}
&\tilde{\bm\gamma}^{i+1}= \\
&{\rm softthreshold}\left(\!\frac{1}{M}\!\!\sum_{m=1}^M\!\!\left(\!\tilde{\bf B}^i_{\rm ER}[:,m]-\frac{1}{\rho_{\rm ER}}{\bf \Upsilon}^i_{\rm ER}[:,m]\!\right),\frac{\zeta_2}{\rho_{\rm ER}}\!\right)\!,
\end{split}\end{equation}
where
\begin{equation}
{\rm softthreshold}(x,a)=\begin{cases}x+a,& x\leq -a,\cr x-a,& x\geq a, \cr 0,& {\rm otherwise}. \end{cases}
\end{equation}
\subsubsection{Update of $\lambda$}
The subproblem with respect to $\lambda$ is:
\begin{equation}
\min_{\lambda}\ \sum_{m=1}^M ||{\bf D}\tilde{\bf B}^i_{\rm ER}[:,m]-\lambda{\bf U}[:,m]-\tilde{\bf Z}_{\rm ER}^{i}[:,m]||_2^2.
\end{equation}
The solution to this subproblem is:
\begin{equation}\label{opti2-ADMM-lmbd}
\lambda^{i+1}\!=\!\max\!\left(\!\!0,\frac{\sum_{m=1}^M{\bf U}[:,m]^T({\bf D}\tilde{\bf B}^i_{\rm ER}[:,m] - \tilde{\bf Z}_{\rm ER}^{i}[:,m])}{||{\bf U}||_2^2}\!\right).
\end{equation}
\subsubsection{Update of $\tilde{\bf B}_{\rm ER}$}
The subproblem with respect to $\tilde{\bf B}_{\rm ER}$ is:
\begin{equation}\begin{split}
&\min_{\tilde{\bf B}_{\rm ER}}\ \sum_{m=1}^M ||{\bf D}\tilde{\bf B}_{\rm ER}[:,m]-\lambda^{i+1}{\bf U}[:,m]-\tilde{\bf Z}_{\rm ER}^{i}[:,m]||_2^2\\
&+\!\!\sum_{m=1}^M\!{\bf\Upsilon}^i_{\rm ER}[:,m]^T(\tilde{\bm\gamma}^{i+1}-\tilde{\bf B}_{\rm ER}[:,m])\!\\
&+\!\frac{\rho_{\rm ER}}{2}\!\!\sum_{m=1}^M\!||\tilde{\bm\gamma}^{i+1}-\tilde{\bf B}_{\rm ER}[:,m]||_2^2.
\end{split}\end{equation}
For given $\{\tilde{\bm\gamma}^{i+1}, \lambda^{i+1}, {\bf\Upsilon}^i_{\rm ER}\}$, the solution to this subproblem is:
\begin{equation}\begin{split}\label{opti2-ADMM-B}
\tilde{\bf B}_{\rm ER} = &({\bf D}^T{\bf D}+\rho_{\rm ER}{\bf I}_{N})^{-1}\\
&(\lambda^{i+1} {\bf D}^T{\bf U}+{\bf D}^T\tilde{\bf Z}_{\rm ER}^{i}+{\bf \Upsilon}^i_{\rm ER}+\rho_{\rm ER}\tilde{\bm\Gamma}^{i+1}), \\
&\tilde{\bm\Gamma} = [\tilde{\bm\gamma},\dots,\tilde{\bm\gamma}]\in\mathbb{R}^{N\times M}.
\end{split}\end{equation}
\subsubsection{Update of ${\bf \Upsilon}_{\rm ER}$}
The Lagrange multiplier ${\bf \Upsilon}_{\rm ER}$ can be updated via gradient descent:
\begin{equation}
\begin{split}\label{opti2-ADMM-U}
{\bf \Upsilon}^{i+1}_{\rm ER}[:,m]\!\leftarrow\!{\bf \Upsilon}^i_{\rm ER}[:,m]+\rho_{\rm ER}(\tilde{\bm\gamma}^{i+1}-\tilde{\bf B}^{i+1}_{\rm ER}[:,m]),\\
m=1,2,\dots,M.
\end{split}
\end{equation}

\subsubsection{Stopping Criteria}
Similar to Section \ref{sec:ADMMSC1}, we stop the ADMM iterations when the primal, dual residuals and the $\lambda$ residual are all sufficiently small. At the $(j+1)$th iteration of ADMM, the three values can be expressed as:
\begin{equation}\label{update_res_ER}\begin{split}
{\rm primal\ residual:}&\ r^{j+1}_{\rm priER} = ||\tilde{\bm\Gamma}^{j+1}-\tilde{\bf B}_{\rm ER}^{j+1}||_2,\\
{\rm dual\ residual:}&\ r^{j+1}_{\rm dual ER} = \rho_{\rm ER}||\tilde{\bf B}_{\rm ER}^{j+1}-\tilde{\bf B}_{\rm ER}^{j}||_2,\\
\lambda\ {\rm residual:}&\ r^{j+1}_{\rm \lambda ER} = ||\lambda^{j+1}-\lambda^j||_1/\lambda^{j+1}.
\end{split}\end{equation}
Then the stopping criteria are:
\begin{equation}\label{sc2}
r^{j}_{\rm priER}\leq \varepsilon_{\rm priER}^{j}, r^{j}_{\rm dual ER}\leq \varepsilon_{\rm dualER}^{j}\ {\rm and}\ r_{\rm \lambda ER}^{j}\leq \varepsilon_{\rm \lambda ER}^{j},
\end{equation}
where $\varepsilon_{\rm \lambda ER}^{j}$ can be set as a small constant and $\varepsilon_{\rm priER}^{j}, \varepsilon_{\rm dualER}^{j}$ can be set as follows:
\begin{equation}\label{update_tol_ER}\begin{split}
&\varepsilon_{\rm priER}^{j} = \sqrt{NM}\varepsilon_{\rm absER} + \varepsilon_{\rm relER}\max (||\tilde{\bm\Gamma}^{j}||_2,||\tilde{\bf B}^{j}_{\rm ER}||_2),\\
&\varepsilon_{\rm dualER}^{j} = \sqrt{NM}\varepsilon_{\rm absER} + \varepsilon_{\rm relER}||{\bf \Upsilon}_{\rm ER}^j||_2,
\end{split}\end{equation}
with $\varepsilon_{\rm absER}$ and $\varepsilon_{\rm relER}$ being absolute and relative tolerances, respectively, which are chosen depending on the scale of the problem.
\subsubsection{Update of $\rho_{\rm ER}$}
Similar to Section \ref{sec:ADMMRHO1}, we update $\rho_{\rm ER}$ by the following scheme:
\begin{equation}\label{update_rho_ER}
\rho_{\rm ER}^{j+1} = \begin{cases}2\rho_{\rm ER}^{j},& r_{\rm priER}^{j} > 10r_{\rm dualER}^{j}, \cr 0.5\rho_{\rm ER}^{j},& r_{\rm dualER}^{j} > r_{\rm priER}^{j}, \cr \rho_{\rm ER}^{j},& {\rm otherwise},\end{cases}
\end{equation}
where $\rho_{\rm ER}^{j}$ denotes the penalty parameter in the $j$th ADMM iteration. The echo recovery algorithm is summarized in Table \ref{alg:ER}.
\begin{table}[!t]
\caption{Echo Recovery}\label{alg:ER}
\centering
\resizebox{0.48\textwidth}{!}{
\begin{tabular}{cl}
\hline
1:&\textbf{Input:} Signed measurement matrix ${\bf Y}$, known threshold matrix ${\bf H}$,\\
& estimated RFI signal $\hat{\bf R}_{\hat{\bm\theta}}$, initial values $\tilde{\bm\gamma}^0$ and $\varepsilon_{\rm absER},\varepsilon_{\rm relER},\varepsilon_{\rm \lambda ER}$.\\
& The initial value of $\lambda$ can be set as the value obtained by 1bMMREALX.\\
2:& ${\bf U} = {\bf H}-\hat{\bf R}_{\hat{\bm\theta}}$.\\
3:&MM Repeat:\\
4:&\quad Initialize $\rho_{\rm ER}$ for ADMM iteration.\\
5:&\quad Construct $\tilde{\bf Z}_{\rm ER}^i$ based on $\tilde{\bf B}_{\rm ER}^i, \lambda^i$ by (\ref{opti2-MM}).\\
6:&\quad ADMM Repeat:\\
7:&\quad\quad Update $\tilde{\bm\gamma}$ by (\ref{opti2-ADMM-A}); \\
8:&\quad\quad Update $\lambda$ by (\ref{opti2-ADMM-lmbd}); \\
9:&\quad\quad Update $\tilde{\bf B}_{\rm ER}$ by (\ref{opti2-ADMM-B}); \\
10:&\quad\quad Update ${\bf \Upsilon}_{\rm ER}$ by (\ref{opti2-ADMM-U}); \\
11:&\quad\quad Calculate $r_{\rm priER},r_{\rm dualER},r_{\rm \lambda ER}, \varepsilon_{\rm priER}, \varepsilon_{\rm dualER}$ by (\ref{update_res_ER}) and (\ref{update_tol_ER});\\
12:&\quad\quad Update $\rho_{\rm ER}$ by (\ref{update_rho_ER}).\\
13:&\quad Until stopping criteria of ADMM iterations $(\ref{sc2})$ are satisfied.\\
14&Until stopping criteria of MM iterations are satisfied.\\
15:&\textbf{Output:} $\hat{\tilde{\bm\gamma}}$ and $\hat{\lambda}$. Recovered Radar Echo $\hat{\bf s} = {\bf D}\hat{\tilde{\bm\gamma}}/\hat{\lambda}$.\\
\hline
\end{tabular}}
\end{table}

\section{Simulated and Experimental Examples}\label{sec:exp}
In this section, we evaluate the RFI mitigation performance of the proposed algorithm for one-bit UWB radar systems using both simulated and measured RFI data sets. Hereafter, the proposed algorithm includes using the fast frequency initialization with the extended 1bMMRELAX for RFI parameter estimation and the extended 1bBIC for RFI order determination (see Section \ref{1bMMRELAX}) and the sparse method to recover the radar echoes (see Section \ref{echo_recover}). We conduct experiments using simulated RFI-free UWB radar data and two different RFI data sets: simulated RFI data set and measured RFI data set. The measured RFI data set was collected by the ARL experimental radar receiver from a real-world environment with the antenna pointing toward Washington DC (see \cite{NTD14, NT16, RNKWS07} for more details about the data collection using the ARL radar). Because the sampling rate of the ARL radar receiver is $8$ GHz, we assume that all data sets used in this section are obtained with an $8$ GHz sampling rate.

\begin{figure}[htbp]
\centering
\subfloat[]{\label{fig:pules}\includegraphics[width=0.48\textwidth]{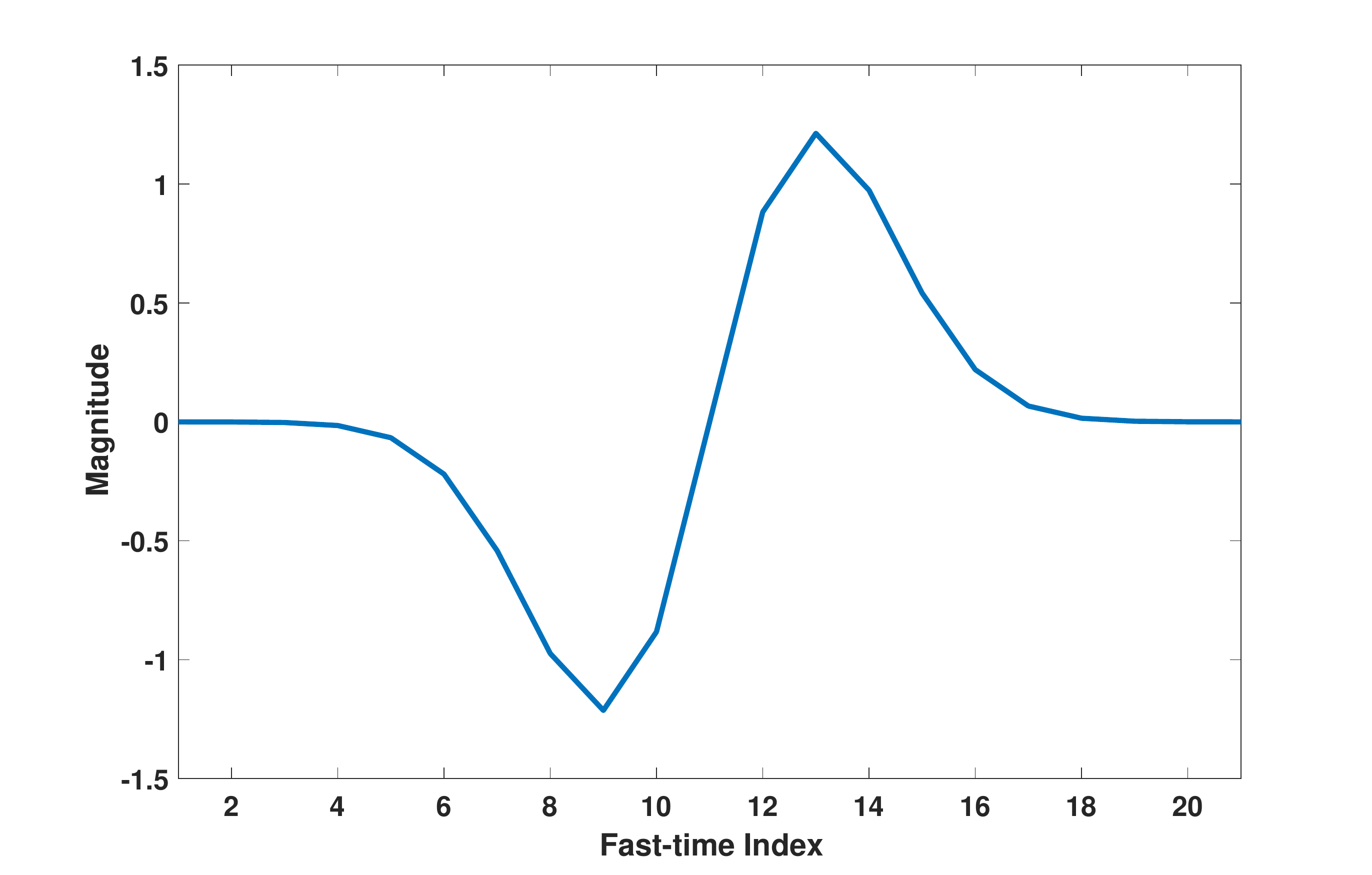}}\\
\subfloat[]{\label{fig:pules_fft}\includegraphics[width=0.48\textwidth]{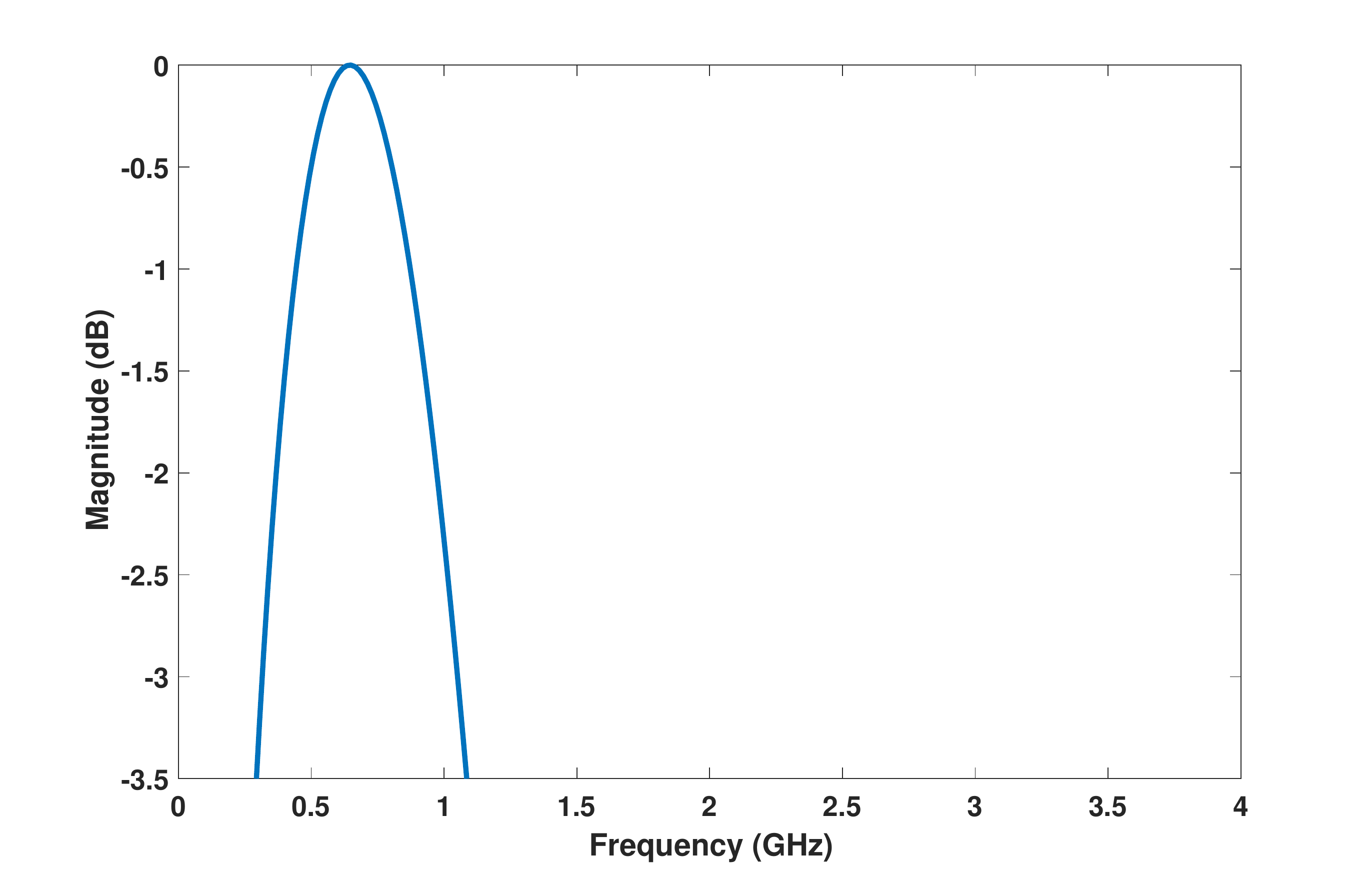}}\\
\subfloat[]{\label{fig:echo}\includegraphics[width=0.48\textwidth]{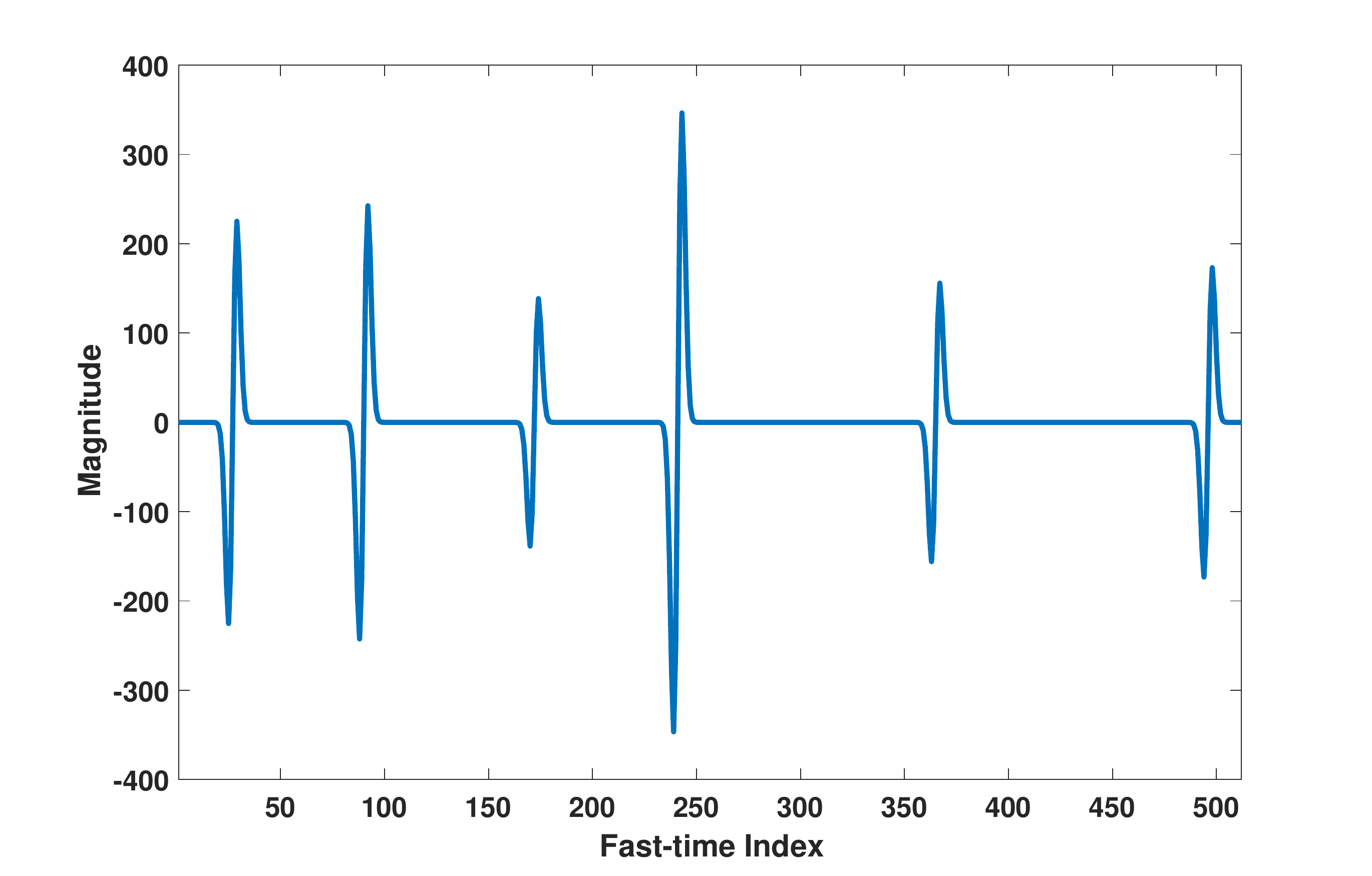}}
\caption{(a) Simulated transmitted radar pulse; (b) Spectrum of the transmitted radar pulse; (c) Simulated RFI-free and noise-free radar echoes for one PRI.}\label{fig:exp_orig}
\end{figure}

All data sets contain $8192$ slow-time samples within a CPI and $512$ fast-time samples per PRI, i.e., $M=8192,N=512$. The transmitted radar impulse is shaped as the first order derivative of a Gaussian pulse with $21$ samples covering the frequency range of $300\sim 1100$ MHz (see Figures \ref{fig:exp_orig}(a) and \ref{fig:exp_orig}(b)). The simulated RFI-free and noise-free UWB radar echoes are generated by $6$ targets at different ranges with different amplitudes (see Figure \ref{fig:exp_orig}(c)).

All signed measurements are obtained by sampling the RFI-contaminated data via the CTBV sampling technology. The DI method described in Section \ref{sec:system} is used as a benchmark. The extended 1bMMRELAX algorithm without the fast frequency initialization is not considered herein because of its computationally prohibitive complexity.

\subsection{Evaluation Metric}
Note that the measured RFI data set inevitably contains noise and other disturbances whereas the simulated RFI data set is noise-free. We add white Gaussian noise ${\bf E}$ to the simulated RFI data set, but we do not add extra noise to the measured (and hence already noisy) RFI data set. Different interference-to-noise ratios (INRs) of the simulated RFI powers to the additive Gaussian noise powers, measured by $20\log_{10}\frac{\|{\bf R}\|_2}{\|{\bf E}\|_2}$ (dB), will be considered in Section \ref{sec:sim_RFI} to show the performance of the proposed algorithm.

Denote the signal-to-interference-plus-noise ratio (SINR) for the simulated RFI data sets and radar echoes as follows:
\begin{equation}
{\rm SINR} = 20\log_{10}\frac{||{\bf S}||_2}{||{\bf R}+{\bf E}||_2}\quad({\rm dB}).
\end{equation}
For the case of simulated radar echoes and measured RFI data sets, with the measured RFI already containing noise, the SINR is computed as follows:
\begin{equation}
{\rm SINR} = 20\log_{10}\frac{||{\bf S}||_2}{||{\bf R}||_2}\quad({\rm dB}).
\end{equation}
We fix the desired radar echo signal and add the scaled simulated RFI plus noise or the scaled measured RFI to obtain contaminated data sets with various SINR values. The maximum threshold $h$ is set as $400$ for all cases according to the magnitude of the desired radar echo signal in Figure \ref{fig:exp_orig}(c).

The radar echo signal recovery performance is measured using the normalized recovery error (NRE) of the recovered signal vector:
\begin{equation}
{\rm NRE} = 20\log_{10}\frac{||{\bf s}-\hat{\bf s}||_2}{||{\bf s}||_2}\quad({\rm dB}),
\end{equation}
where $\hat{\bf s}$ is the recovered UWB radar echo signal.
\subsection{Implementation Details}
In our implementation of 1bMMRELAX, the MM iterations are terminated if the relative change of the negative log-likelihood function $l({\tilde{\bm\beta}})$ between two consecutive iterations is less than $10^{-7}$ or a maximum number of the MM iterations $T_M = 20$ is reached. Within each MM iteration, we terminate the inner loop if the relative change of the objective function in (\ref{MM}) is less than $10^{-9}$ or a maximum number of iterations $T_C=20$ is reached. When using the $N_1$-point FFT in the infinite-precision RELAX, $N_1$ is set to $64N$. When using the fast frequency initialization, we set $Q = N, \varepsilon_{\rm absFI}=10^{-3}, \varepsilon_{\rm relFI}=10^{-7}$ and $\varepsilon_{\rm \lambda FI} = 10^{-3}$, and set the user-parameter $\zeta_1=1$ for all cases. We terminate the inner loop, i.e., the ADMM iteration, when the stopping criteria in (\ref{sc1}) is satisfied or the iteration number is larger than 10. The outer loop, i.e., the MM iteration, will be terminated if the relatively change of the values of the objective function (\ref{opti1}) is smaller than $10^{-7}$ or the iteration number is larger than 5. In the final radar echo signal recovery step, we set $\varepsilon_{\rm absER}=10^{-3}, \varepsilon_{\rm relER}=10^{-7}$ and $\varepsilon_{\rm \lambda ER} = 10^{-3}$, and $\zeta_2=0.04$ for all cases. We terminate the inner loop, i.e., the ADMM iteration, when the stopping criteria in (\ref{sc2}) is satisfied or the iteration number is larger than 100. The outer loop, i.e., the MM iteration, will be terminated if the relative change of the values of the objective function (\ref{opti2}) is smaller than $10^{-7}$ or the iteration number is larger than 50.

\subsection{Simulated RFI cases}\label{sec:sim_RFI}
\begin{figure}[htbp]
\centering
\includegraphics[width=0.48\textwidth]{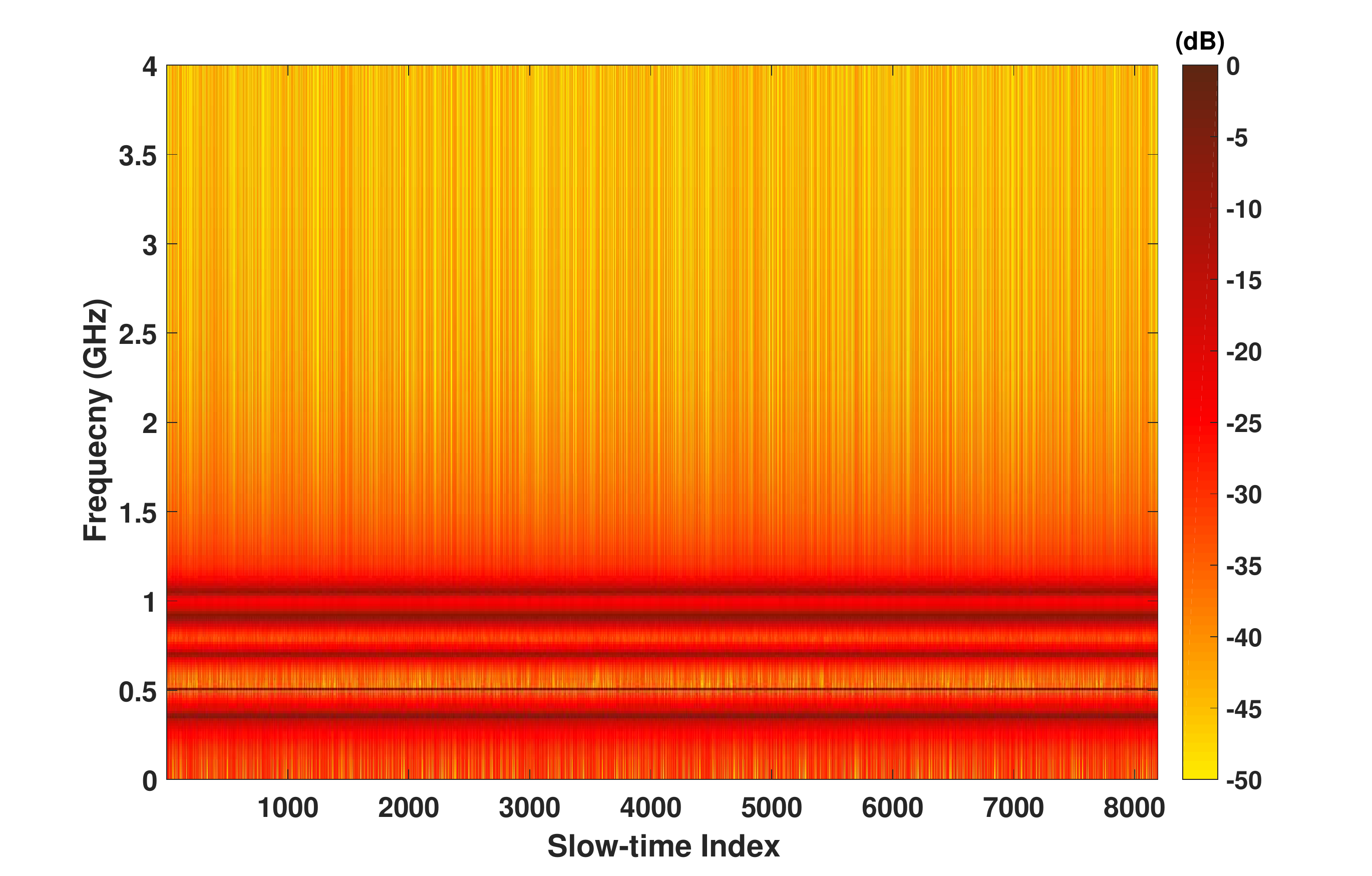}
\caption{Spectrum of the simulated RFI.}
\label{fig:spec_RFI_sim}
\end{figure}

In this section, we present the results of the proposed algorithm for the simulated radar echoes contaminated by the simulated RFI. The magnitudes of the RFI sources usually do not change greatly among different PRIs within a CPI. Thus we simulate the RFI sources as a sum of sinusoids with amplitudes and frequencies fixed from one PRI to another within a CPI and the phases varying independently (with uniform distribution between $0$ and $2\pi$) with slow-time. The simulated RFI can be written as follows:
\begin{equation}\begin{split}
{\bf R}_{\rm simulated}[n,m] &= \sum_{k=1}^K A_k \sin(\omega_k(n-1)+\phi_{k,m})\\
\end{split}\end{equation}
where $\frac{A_k}{A_1}, k=2,\dots,K$, is the amplitude ratio of the $k$-th RFI source relative to the first one. The detailed parameter settings of the simulated RFI data set are shown in Table \ref{tab:sim_RFI} and Figure \ref{fig:spec_RFI_sim}. When generating the contaminated data sets with different SINR values, the desired RFI can be obtained by varying $A_1$ while fixing $\{\frac{A_k}{A_1}\}_{k=2}^K$.

\begin{table}[htbp]
\caption{Simulated RFI Parameter Settings}\label{tab:sim_RFI}
\centering
\resizebox{0.48\textwidth}{!}{
\begin{tabular}{|c|c|c|c|c|c|}
\hline
RFI Frequencies (MHz)& $500$ & $350$ & $700$ & $900$ & $1050$ \\
\hline
RFI Amplitude Ratios & $1$&$0.95$&$0.8$&$0.87$&$0.9$ \\
\hline
\end{tabular}}
\end{table}

\begin{figure}[htbp]
\centering
\subfloat[]{\includegraphics[width=0.48\textwidth, height = 0.3\textwidth]{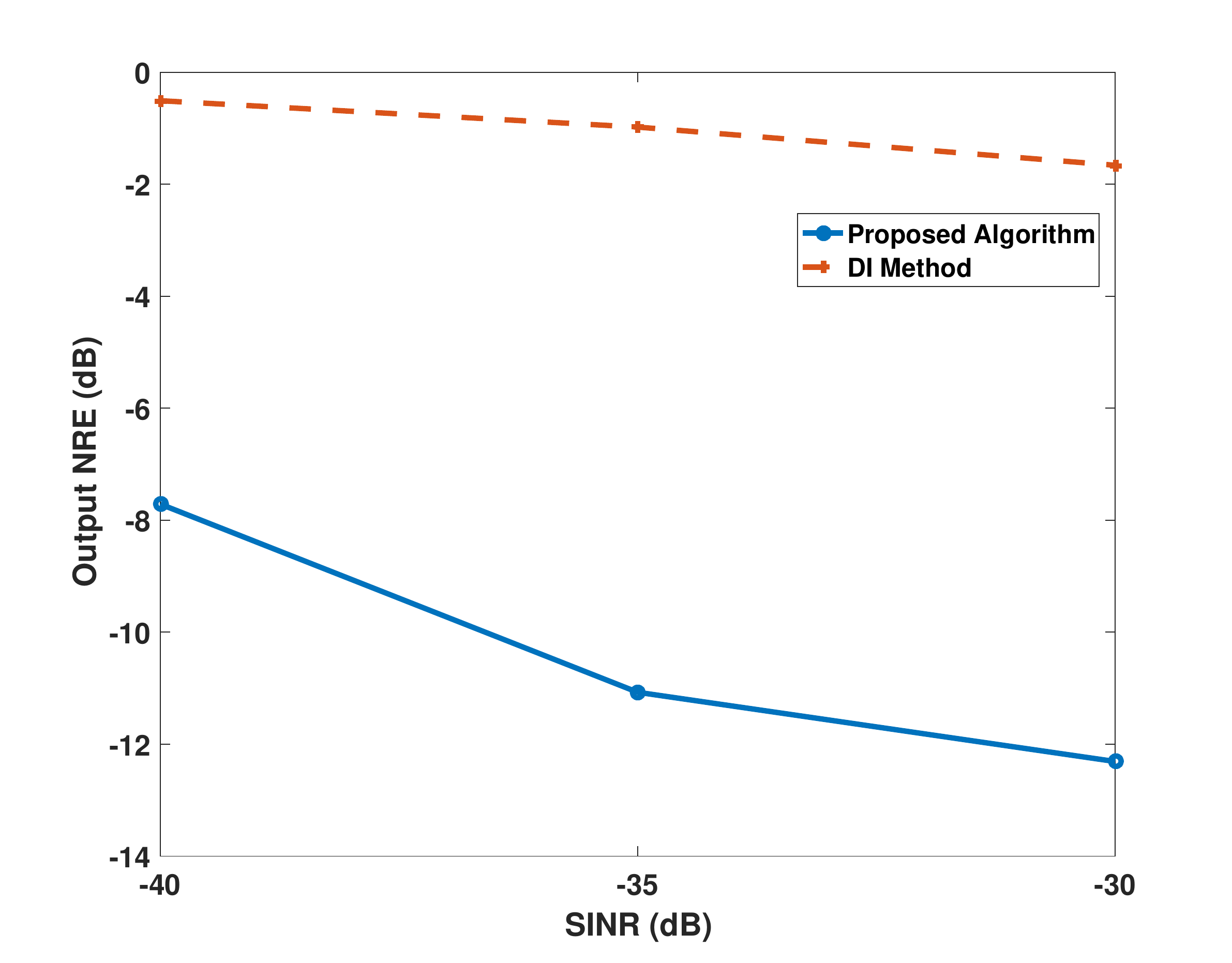}}\\
\subfloat[]{\includegraphics[width=0.48\textwidth, height = 0.3\textwidth]{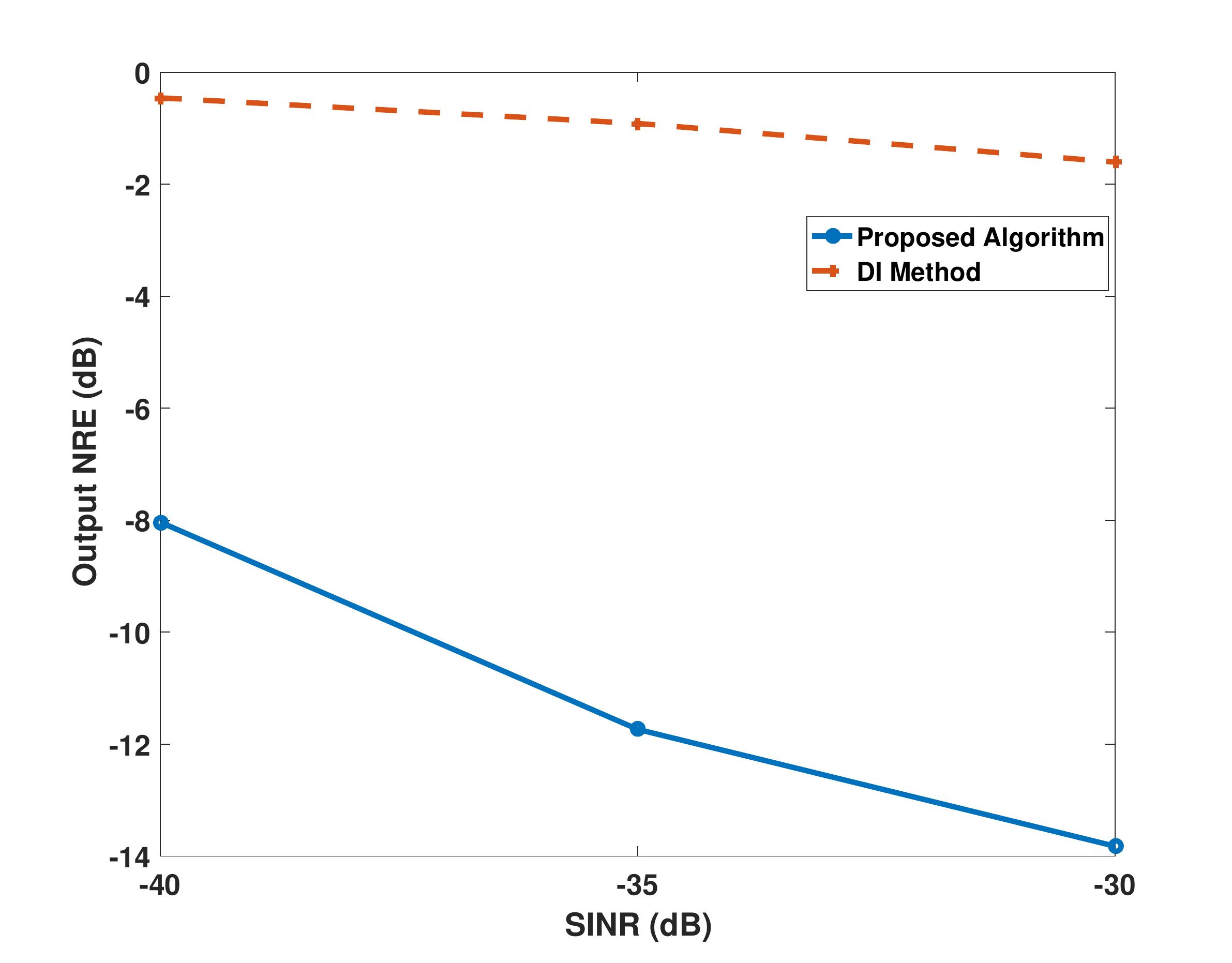}}
\caption{Results of the proposed method and the DI method for simulated RFI data sets with fixed $\zeta_2 = 0.04$. The INR is a) $0$ dB and b) $10$ dB.}
\label{fig:simRFI_fixZ2}
\end{figure}

\begin{figure}[htbp]
\centering
\subfloat[]{\includegraphics[width=0.48\textwidth, height = 0.3\textwidth]{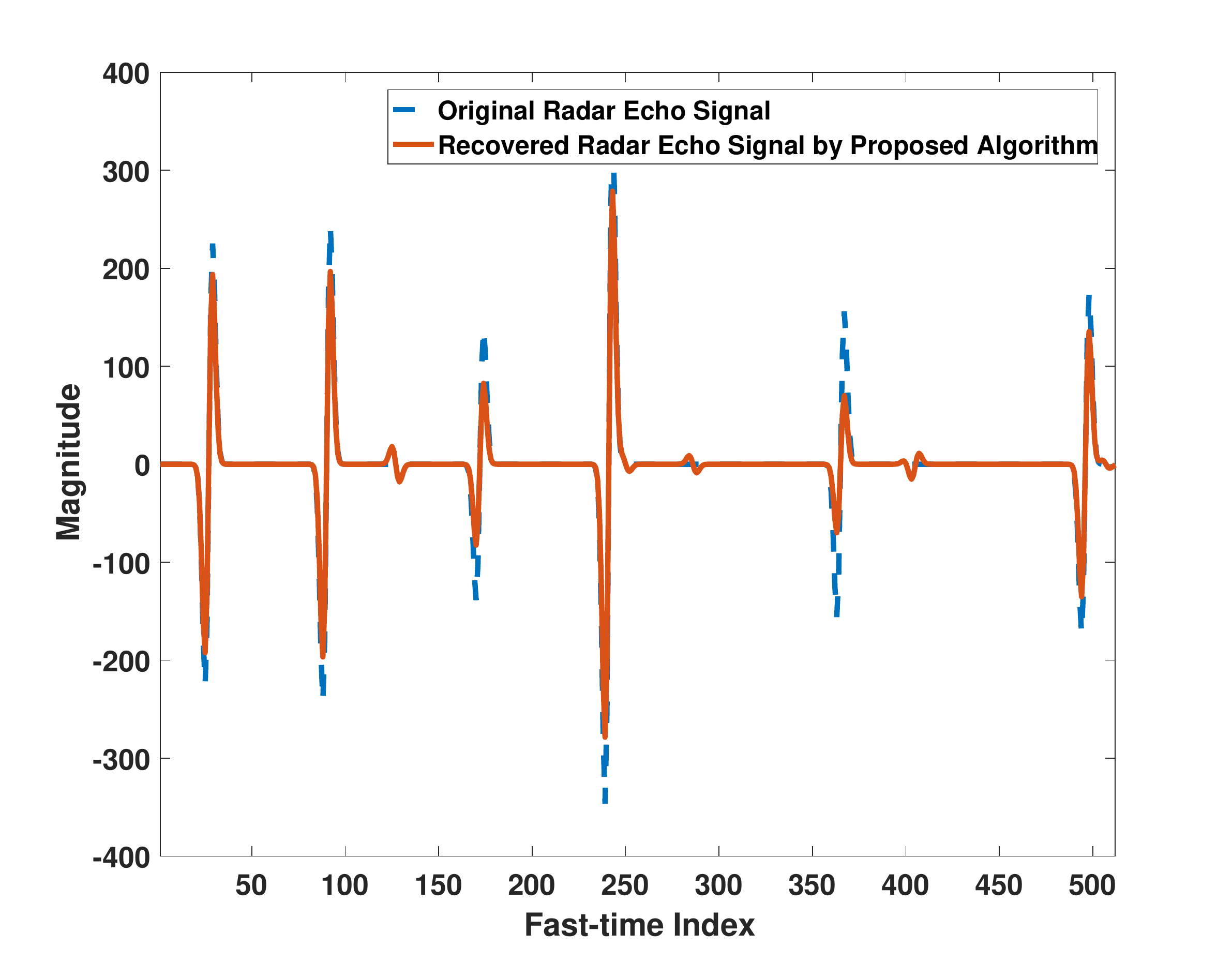}}\\
\subfloat[]{\includegraphics[width=0.48\textwidth, height = 0.3\textwidth]{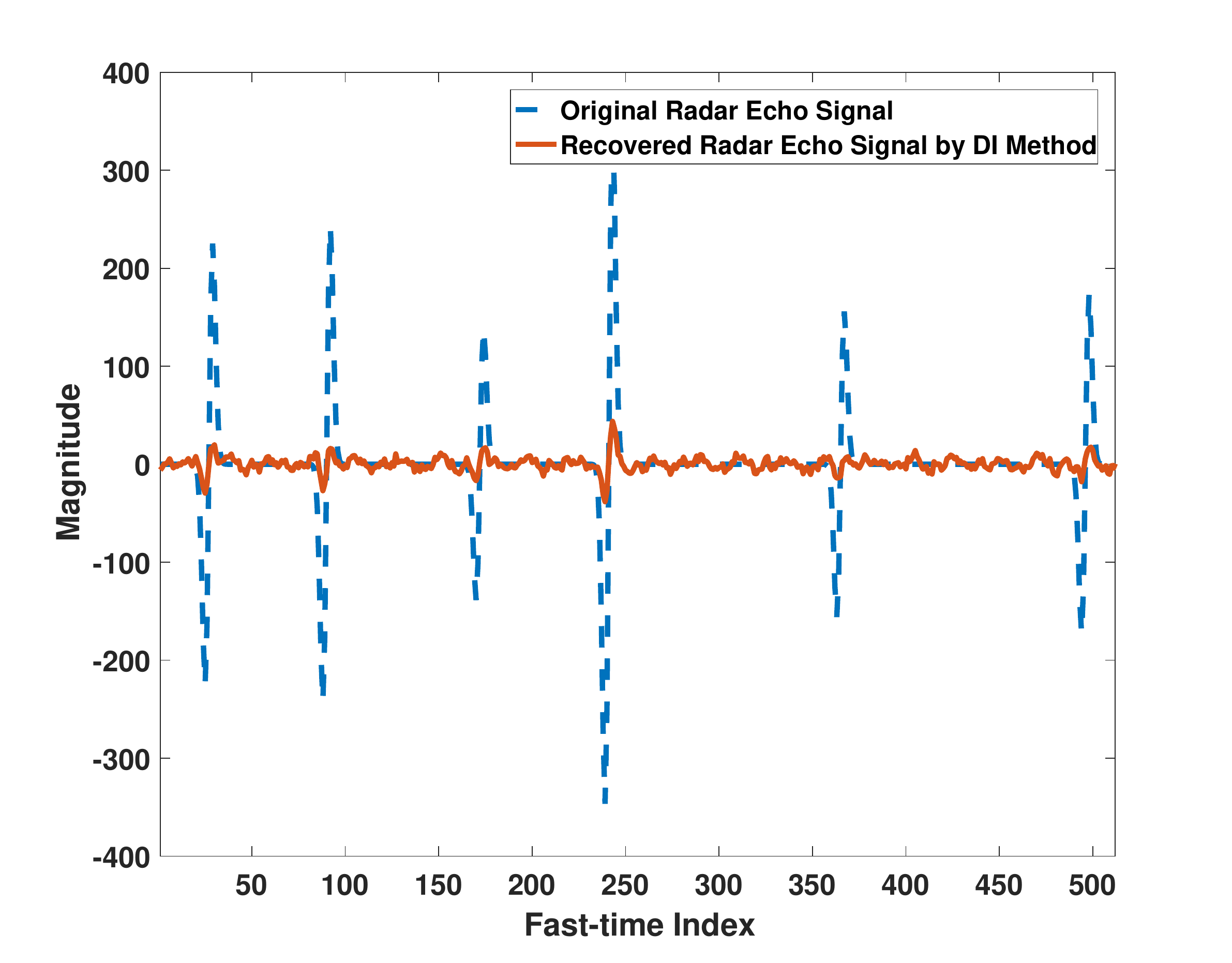}}
\caption{Radar echo recovery results for the simulated RFI obtained by using a) the proposed algorithm and b) the DI method, when SINR = $-35$ dB and INR $=10$ dB.}
\label{fig:simRFI_fixZ2_35_RN10}
\end{figure}

\begin{figure}[htbp]
\centering
\subfloat[]{\includegraphics[width=0.48\textwidth, height = 0.3\textwidth]{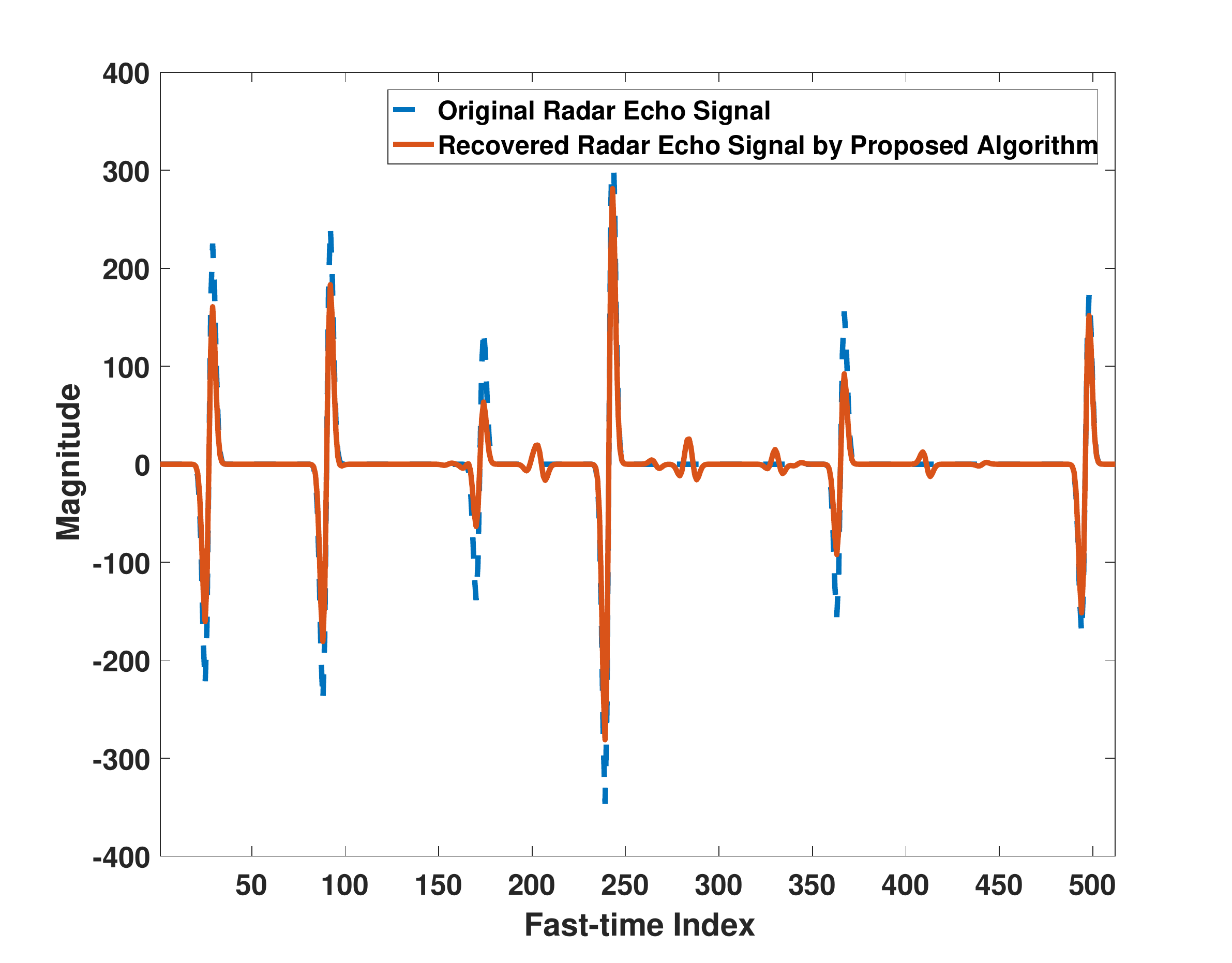}}\\
\subfloat[]{\includegraphics[width=0.48\textwidth, height = 0.3\textwidth]{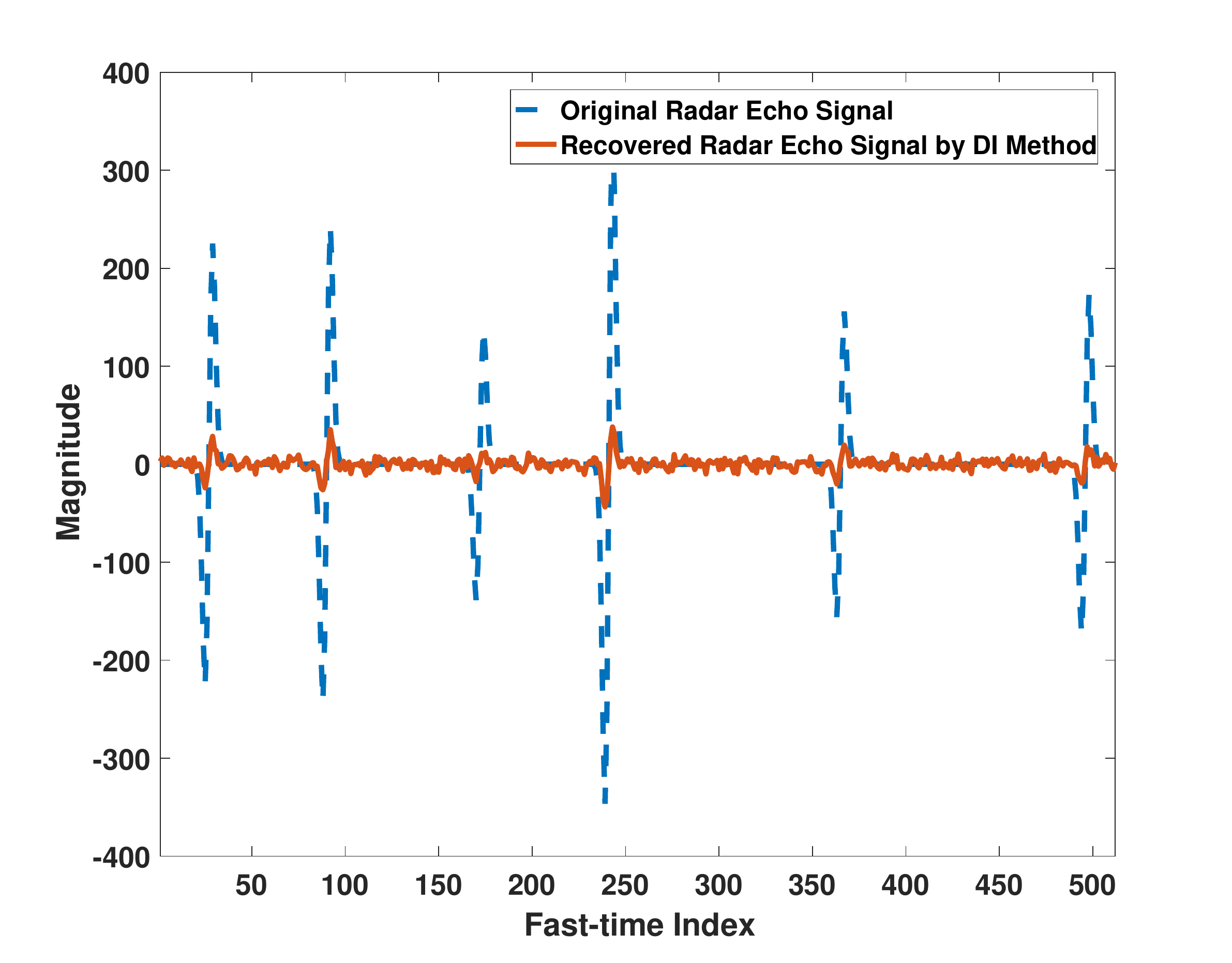}}
\caption{Radar echo recovery results for the simulated RFI obtained by using a) the proposed algorithm and b) the DI method, when SINR = $-35$ dB and INR $=0$ dB.}
\label{fig:simRFI_fixZ2_35_RN0}
\end{figure}

The RFI mitigation results for the simulated RFI data set obtained by the proposed algorithm and the DI method are shown in Figures \ref{fig:simRFI_fixZ2} - \ref{fig:simRFI_fixZ2_35_RN0}. It is clear that the proposed algorithm outperforms the DI method as SINR varies from $-30$ dB to $-40$ dB. Additionally, for the challenging cases of SINR $=-35$ dB, for example, most of the targets are retained by the proposed algorithm, while few targets are distinguishable by the DI method. The results also show that using the extended 1bMMRELAX with the extended 1bBIC provides accurate model order estimates for the multiple-PRI based signed measurements for a wide range of SINR values. Note that the choices of the user parameters $\zeta_1$ and $\zeta_2$ work well even as SINR varies from $-30$ dB to $-40$ dB.

\subsection{Measured RFI cases}
In this section, we present RFI mitigation results using the aforementioned simulated RFI-free UWB radar echoes and the measured RFI data set collected by the ARL experimental radar receiver \cite{NTD14, NT16}. We use the first $512\times 8192$ subset of the original RFI data set measured by the ARL radar receiver. The spectrum of the measured RFI data set is shown in Figure \ref{RADAR2}.

\begin{figure}[htbp]
\centering
\includegraphics[width=0.48\textwidth, height = 0.3\textwidth]{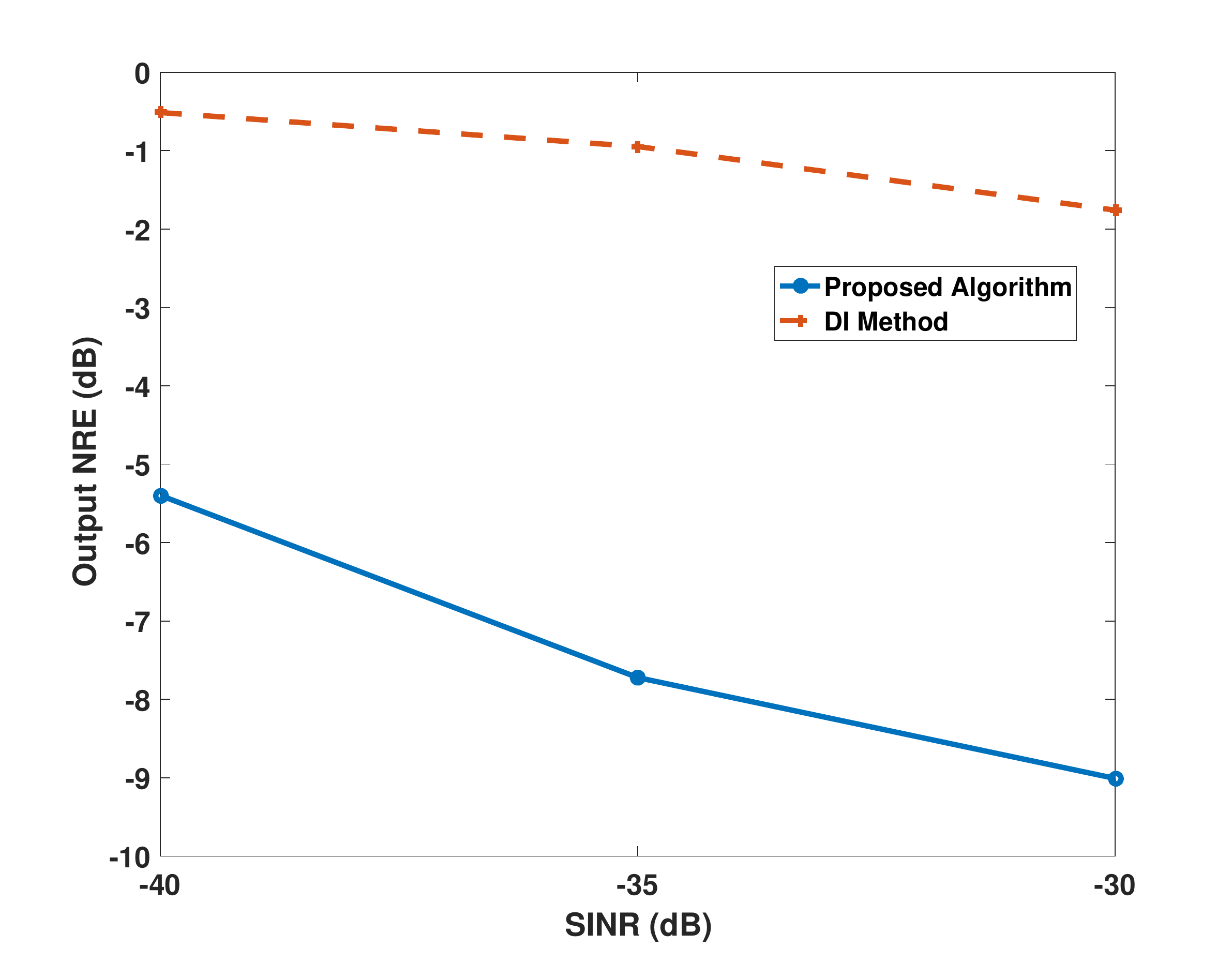}
\caption{Results of the proposed algorithm and the DI method for the measured RFI data sets.}
\label{fig:meaRFI_SINR}
\end{figure}

\begin{figure}[htbp]
\centering
\subfloat[]{\includegraphics[width=0.48\textwidth, height = 0.3\textwidth]{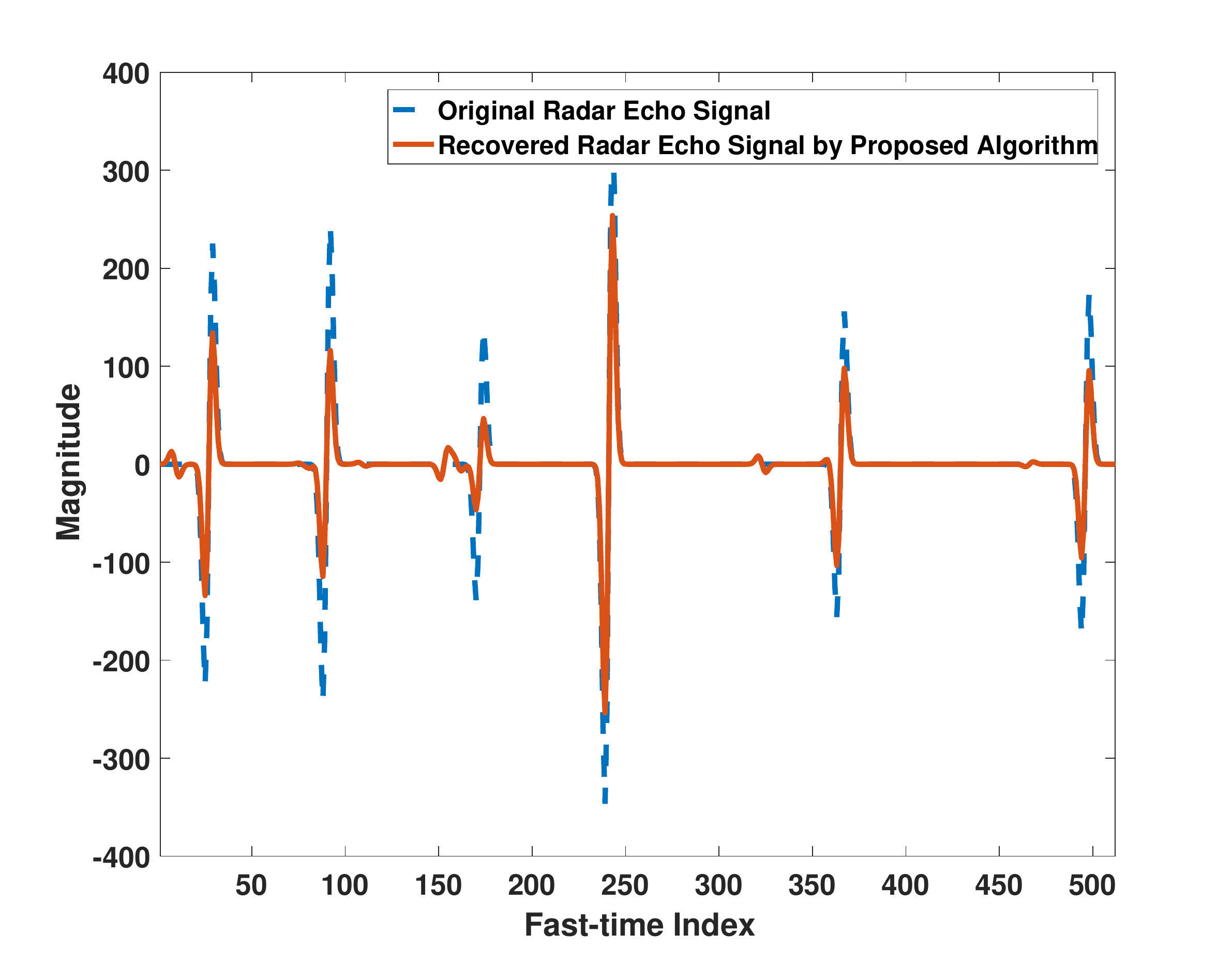}}\\
\subfloat[]{\includegraphics[width=0.48\textwidth, height = 0.3\textwidth]{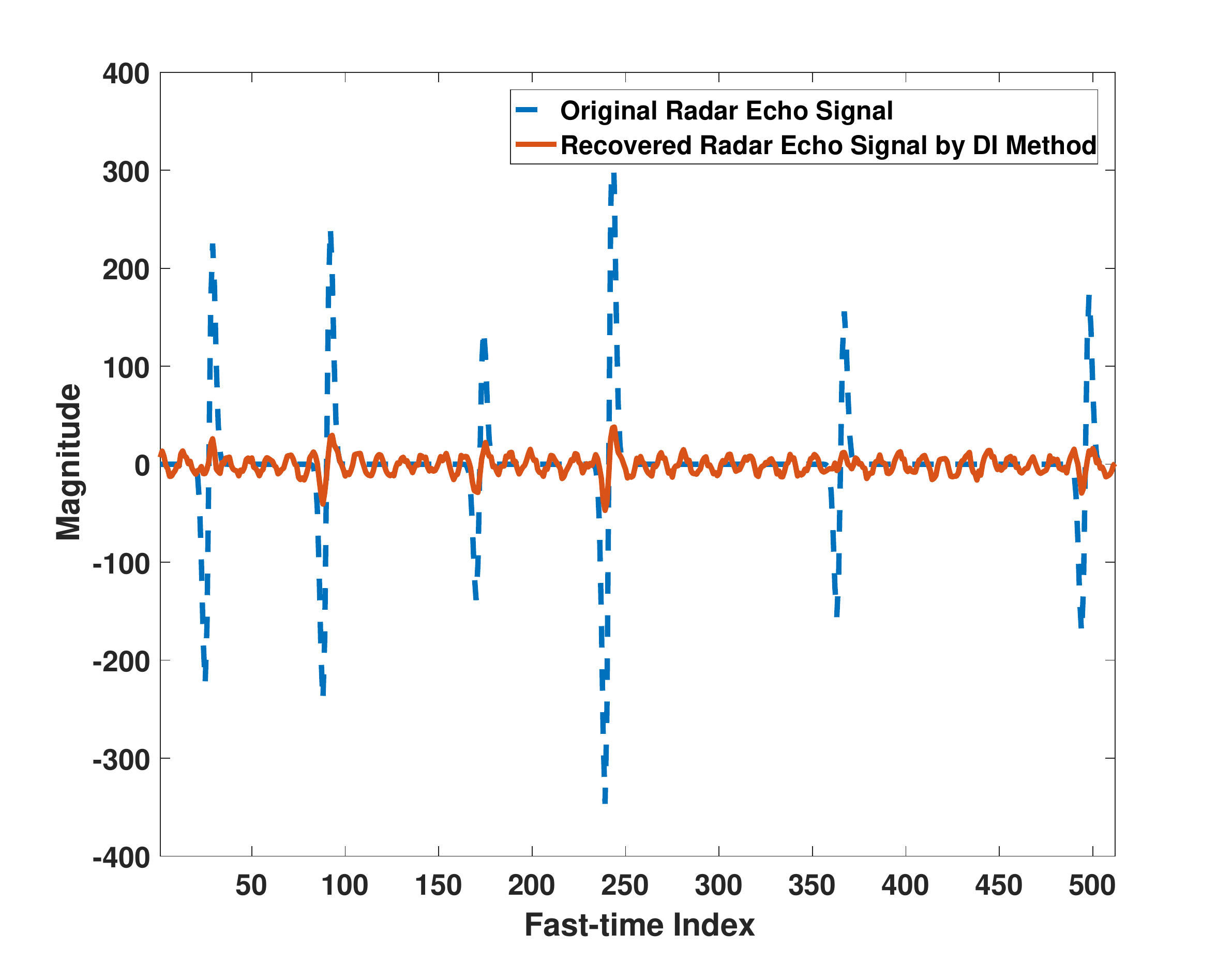}}
\caption{Radar echo recovery results for the measured RFI obtained by using a) the proposed algorithm and b) the DI method, when SINR = $-35$ dB.}
\label{fig:meaRFI_fixZ2_35}
\end{figure}

The RFI mitigation results for the measured RFI data set by the proposed algorithm and the DI method are shown in Figures \ref{fig:meaRFI_SINR} - \ref{fig:meaRFI_fixZ2_35}. Over a wide range of SINR values, the proposed algorithm outperforms the DI method. More specifically, for the challenging case of SINR $=-35$ dB, most of the targets are retained by the proposed algorithm, while few targets are discernible by the DI method.

\section{Conclusions}
In this paper, we have established a RFI mitigation framework for a one-bit UWB radar system that obtains its signed measurements by using the CTBV sampling technique. We have established a proper data model for the RFI sources and have extended a majorization-minimization based 1bMMRELAX algorithm to efficiently estimate the RFI parameters from multiple-PRI based signed measurements. We have also introduced a fast frequency initialization algorithm, based on the MM and ADMM techniques, to further reduce the computational cost of the extended 1bMMRELAX. Moreover, we have extended 1bBIC to the multiple-PRI based cases so that the extended 1bMMRELAX can be used with the extended 1bBIC to simultaneously estimate the RFI parameters and determine the number of RFI sources. Next, by exploiting the sparse property of the UWB radar echoes, a sparse signal recovery method, also based on the MM and ADMM techniques, has been devised to estimate the desired UWB radar echoes using the estimated RFI parameters. Finally, we have provided examples of using both simulated and measured RFI data sets to demonstrate that the proposed algorithm can outperform the benchmark DI method significantly for radar echo recovery, especially for the severe RFI cases.


%

\appendices
\section{Derivation of 1bBIC for Multiple-PRI Based Signed Measurements}
Rewrite the signed measurement matrix ${\bf Y}$ as follows:
\begin{equation}\begin{split}
{\bf Y} &= {\rm sign}({\bf R}_{\breve{\bm\theta}}+{\bf S}+{\bf E}-{\bf H})\in\mathbb{R}^{N\times M},
\end{split}\end{equation}
where
\begin{equation}\begin{split}
{\bf R}_{\breve{\bm\theta}}[n,m] &= \sum_{k=1}^K A_{k,m}\sin(\omega_k(n-1)+\phi_{k,m}),\\
\end{split}\end{equation}
with $\breve{\bm\theta}=[\omega_1,\dots,\omega_K,A_{1,1},\dots,A_{K,M},\phi_{1,1},\dots,\phi_{K,M}]^T\in\mathbb{R}^{(2M+1)K}$ denoting the unknown parameter vector of the RFI sources. Assuming that ${\bf S}+{\bf E}$ obeys i.i.d. Gaussian distribution with zero-mean and unknown variance $\sigma^2$, the unknown parameter vector of ${\bf Y}$ is denoted by $\breve{\bm\beta}=[\breve{\bm\theta},\sigma]^T$.

By assuming that the prior probability density function (PDF) of $\breve{\bm\beta}$ is flat around the ML estimate and independent of $N$ and $M$, the estimate of the signal model order $K$ can be obtained by minimizing the following criterion \cite{SS04-2}:
\begin{equation}\label{bic}
2l_{\bf Y}(\hat{\breve{\bm\beta}})+\log \det(\hat{\bf J}({\bf Y},\hat{\breve{\bm\beta}})),
\end{equation}
where $l_{\bf Y}(\hat{\breve{\bm\beta}})$ denotes the negative log-likelihood function of ${\bf Y}$ and $\hat{\breve{\bm\beta}}$ is the ML estimate of $\breve{\bm\beta}$. With the aforementioned Gaussian distribution assumption, $l_{\bf Y}(\hat{\breve{\bm\beta}})$ can be expressed as follows:
\begin{equation}
l_{\rm Y}(\hat{\breve{\bm\beta}}) = -\!\!\sum_{m=1}^M\!\sum_{n=1}^N\log\!\!\left[\!\Phi\!\left(\!{\bf Y}[n,m]\frac{{\bf R}_{\breve{\bm\theta}}[n,m]\!-\!{\bf H}[n,m]}{\sigma}\!\right)\!\right].
\end{equation}
The function $\det(\cdot)$ means taking the determinant of a matrix. The matrix $\hat{\bf J}({\bf Y},\hat{\breve{\bm\beta}})$ in (\ref{bic}) is defined as:
\begin{equation}
\hat{\bf J}({\bf Y},\hat{\breve{\bm\beta}}) = \frac{\partial^2l_{\bf Y}(\hat{\breve{\bm\beta}})}{\partial\breve{\bm\beta}\partial\breve{\bm\beta}^T}\Bigg|_{\breve{\bm\beta}=\hat{\breve{\bm\beta}}}.
\end{equation}
Under mild conditions (see, e.g., \cite{LZLS18,SS04-2} and the references therein), the matrix $\hat{\bf J}({\bf Y},\hat{\breve{\bm\beta}})$ has the following asymptotic relationship with the Fisher information matrix (FIM) $\bf J$ for the parameter vector $\breve{\bm\beta}$ \cite{LZLS18}:
\begin{equation}\label{J-hatJ}
[{\bf P}_N{\bf J}{\bf P}_N-{\bf P}_N\hat{\bf J}({\bf Y},\hat{\breve{\bm\beta}}){\bf P}_N]\rightarrow 0, {\rm as}\ N\rightarrow\infty,
\end{equation}
where ${\bf P}_N$ is a normalization matrix, and
\begin{equation}
{\bf J} = E\left\{\frac{\partial^2l_{\bf Y}(\hat{\breve{\bm\beta}})}{\partial\breve{\bm\beta}\partial\breve{\bm\beta}^T}\right\}=E\left\{\frac{\partial l_{\bf Y}(\hat{\breve{\bm\beta}})}{\partial\breve{\bm\beta}}\frac{\partial l_{\bf Y}(\hat{\breve{\bm\beta}})}{\partial\breve{\bm\beta}^T}\right\}.
\end{equation}
Hence, after proper normalizations, $\hat{\bf J}({\bf Y},\hat{\breve{\bm\beta}})$ in (\ref{bic}) can be substituted by the FIM in the large sample (i.e., $N\gg1$) case, a fact that can be used to obtain a simple asymptotically valid expression for the penalty term in (\ref{bic}).

Let ${\bf u}_{n,m}=[\frac{\partial{\bf R}_{\breve{\bm\theta}}[n,m]}{\partial\breve{\bm\theta}^T},-\frac{{\bf R}_{\breve{\bm\theta}}[n,m]-{\bf H}[n,m]}{\sigma}]^T$. After simple calculations, the FIM can be expressed as a sum of positive semidefinite matrices \cite{GXLS16}:
\begin{equation}
{\bf J} = \frac{1}{2\pi\sigma^2}\sum_{n=1}^N\sum_{m=1}^M\xi\left(\frac{{\bf R}_{\breve{\bm\theta}}[n,m]-{\bf H}[n,m]}{\sigma}\right){\bf u}_{n,m}{\bf u}_{n,m}^T,
\end{equation}
where $\xi(x)$ is defined as follows:
\begin{equation}
\xi(x)=\left[\frac{1}{\Phi(x)}+\frac{1}{\Phi(-x)}\right]e^{-x^2}.
\end{equation}
Let ${\bf J}_1=\frac{1}{\sigma^2}\sum_{n=1}^N\sum_{m=1}^M{\bf u}_{n,m}{\bf u}_{n,m}^T$. Note that $\log\det({\bf J})$ and $\log\det({\bf J}_1)$ are asymptotically equivalent (see \cite{LZLS18} for a detailed proof), to within a constant that does not depend on $N$. Therefore, the FIM ${\bf J}$ can be substituted by the simpler matrix ${\bf J}_1$. Note that ${\bf J}_1$ is not equal to the FIM for the conventional BIC. Next, we analyze ${\bf J}_1$ and the normalization matrix ${\bf P}_N$ needed for ${\bf P}_N{\bf J}_1{\bf P}_N$ to be $O(1)$ for our RFI signal model.

Under the weak assumption that the threshold matrix ${\bf H}$ is uncorrelated with the RFI ${\bf R}_{\breve{\bm\theta}}$, we have the following results (see \cite{LZLS18} and Appendix A of \cite{SMFS98}):
\begin{equation}\begin{split}\label{omega-omega}
&\lim_{N\rightarrow\infty}\frac{1}{\sigma^2N^3}\sum_{n=1}^N\sum_{m=1}^M\frac{\partial{\bf R}_{\breve{\bm\theta}}[n,m]}{\partial\omega_{k_i}}\frac{\partial{\bf R}_{\breve{\bm\theta}}[n,m]}{\partial\omega_{k_j}}\\
&=\lim_{N\rightarrow\infty}\frac{1}{\sigma^2N^3}\sum_{n=1}^N\sum_{m=1}^M
(n-1)^2A_{k_i,m}A_{k_j,m}\\
&\cos(\omega_{k_i}(n-1)+\phi_{k_i,m})\cos(\omega_{k_j}(n-1)+\phi_{k_j,m})\\
&=\frac{\delta_{i,j}}{6\sigma^2}\sum_{m=1}^MA_{k_i,m}^2,\\
&\delta_{i,j} = \begin{cases}0, &i\neq j \cr 1,&i=j\end{cases}\quad 1\leq i,j\leq K,
\end{split}\end{equation}

\begin{equation}\begin{split}
&\lim_{N\rightarrow\infty}\frac{1}{\sigma^2N^2}\sum_{n=1}^N\sum_{m=1}^M\frac{\partial{\bf R}_{\breve{\bm\theta}}[n,m]}{\partial\omega_{k_i}}\frac{\partial{\bf R}_{\breve{\bm\theta}}[n,m]}{\partial A_{k_j,m'}}\\
&=\lim_{N\rightarrow\infty}\frac{1}{\sigma^2N^2}\sum_{n=1}^N(n-1)A_{k_i,m'}\\
&\cos(\omega_{k_i}(n-1)+\phi_{k_i,m'})\sin(\omega_{k_j}(n-1)+\phi_{k_j,m'})\\
&=0,\ 1\leq m' \leq M,
\end{split}\end{equation}

\begin{equation}\begin{split}
&\lim_{N\rightarrow\infty}\frac{1}{\sigma^2N^2}\sum_{n=1}^N\sum_{m=1}^M\frac{\partial{\bf R}_{\breve{\bm\theta}}[n,m]}{\partial\omega_{k_i}}\frac{\partial{\bf R}_{\breve{\bm\theta}}[n,m]}{\partial \phi_{k_j,m'}}\\
&=\lim_{N\rightarrow\infty}\frac{1}{\sigma^2N^2}\sum_{n=1}^N(n-1)A_{k_i,m'}A_{k_j,m'}\\
&\cos(\omega_{k_i}(n-1)+\phi_{k_i,m'})\cos(\omega_{k_j}(n-1)+\phi_{k_j,m'})\\
&=\frac{\delta_{i,j}}{4\sigma^2}A_{k_i,m'}^2,
\end{split}\end{equation}

\begin{equation}\begin{split}
&\lim_{N\rightarrow\infty}\frac{-1}{\sigma^2N^2}\sum_{n=1}^N\sum_{m=1}^M\frac{\partial{\bf R}_{\breve{\bm\theta}}[n,m]}{\partial\omega_{k_i}}\frac{{\bf R}_{\breve{\bm\theta}}[n,m]-{\bf H}[n,m]}{\sigma}\\
&=\lim_{N\rightarrow\infty}\frac{-1}{\sigma^3N^2}\sum_{n=1}^N\sum_{m=1}^M(n-1)A_{k_i,m}\\
&\cos(\omega_{k_i}(n-1)+\phi_{k_i,m})({\bf R}_{\breve{\bm\theta}}[n,m]-{\bf H}[n,m])\\
&=0,
\end{split}\end{equation}

\begin{equation}\begin{split}
&\lim_{N\rightarrow\infty}\frac{1}{\sigma^2N}\sum_{n=1}^N\sum_{m=1}^M\frac{\partial{\bf R}_{\breve{\bm\theta}}[n,m]}{\partial A_{k_i,m_p}}\frac{\partial{\bf R}_{\breve{\bm\theta}}[n,m]}{\partial A_{k_j,m_q}}, 1\leq p, q\leq M\\
&=\lim_{N\rightarrow\infty}\frac{1}{\sigma^2N}\sum_{n=1}^N\sin(\omega_{k_i}(n-1)+\phi_{k_i,m_p})\\
&\sin(\omega_{k_j}(n-1)+\phi_{k_j,m_p})\delta_{p,q}\\
&=\frac{1}{2\sigma^2}\delta_{i,j}\delta_{p,q},
\end{split}\end{equation}

\begin{equation}\begin{split}
&\lim_{N\rightarrow\infty}\frac{1}{\sigma^2N}\sum_{n=1}^N\sum_{m=1}^M\frac{\partial{\bf R}_{\breve{\bm\theta}}[n,m]}{\partial A_{k_i,m_p}}\frac{\partial{\bf R}_{\breve{\bm\theta}}[n,m]}{\partial \phi_{k_j,m_q}}\\
&=\lim_{N\rightarrow\infty}\frac{1}{\sigma^2N^2}\sum_{n=1}^NA_{k_j,m_p}\sin(\omega_{k_i}(n-1)+\phi_{k_i,m_p})\\
&\cos(\omega_{k_j}(n-1)+\phi_{k_j,m_p})\delta_{p,q}\\
&=0,
\end{split}\end{equation}

\begin{equation}\begin{split}
&\lim_{N\rightarrow\infty}\frac{-1}{\sigma^2N}\sum_{n=1}^N\sum_{m=1}^M\frac{\partial{\bf R}_{\breve{\bm\theta}}[n,m]}{\partial A_{k_i,m'}}\frac{{\bf R}_{\breve{\bm\theta}}[n,m]-{\bf H}[n,m]}{\sigma}\\
&=\lim_{N\rightarrow\infty}\frac{-1}{\sigma^3N}\sum_{n=1}^N\sin(\omega_{k_i}(n-1)+\phi_{k_i,m'})\\
&({\bf R}_{\breve{\bm\theta}}[n,m']-{\bf H}[n,m'])\\
&=\lim_{N\rightarrow\infty}\frac{-1}{\sigma^3N}\sum_{n=1}^N\Bigg[\sin(\omega_{k_i}(n-1)+\phi_{k_i,m'})\\
&\sum_{k=1}^K A_{k,m'}\sin(\omega_k(n-1)+\phi_{k,m'})\Bigg]\\
&=\frac{-A_{k_i,m'}}{2\sigma^3},
\end{split}\end{equation}

\begin{equation}\begin{split}
&\lim_{N\rightarrow\infty}\frac{1}{\sigma^2N}\sum_{n=1}^N\sum_{m=1}^M\frac{\partial{\bf R}_{\breve{\bm\theta}}[n,m]}{\partial \phi_{k_i,m_p}}\frac{\partial{\bf R}_{\breve{\bm\theta}}[n,m]}{\partial \phi_{k_j,m_q}}\\
&=\lim_{N\rightarrow\infty}\frac{1}{\sigma^2N}\sum_{n=1}^N A_{k_i,m_p}A_{k_j,m_p}\cos(\omega_{k_i}(n-1)+\phi_{k_i,m_p})\\
&\cos(\omega_{k_j}(n-1)+\phi_{k_j,m_p})\delta_{p,q}\\
&=\frac{A_{k_i,m_p}^2}{2\sigma^2}\delta_{i,j}\delta_{p,q},
\end{split}\end{equation}

\begin{equation}\begin{split}
&\lim_{N\rightarrow\infty}\frac{-1}{\sigma^2N}\sum_{n=1}^N\sum_{m=1}^M\frac{\partial{\bf R}_{\breve{\bm\theta}}[n,m]}{\partial\phi_{k_i,m'}}\frac{{\bf R}_{\breve{\bm\theta}}[n,m]-{\bf H}[n,m]}{\sigma}\\
&=\lim_{N\rightarrow\infty}\frac{-1}{\sigma^3N}\sum_{n=1}^NA_{k_i,m'}\cos(\omega_{k_i}(n-1)+\phi_{k_i,m'})\\
&({\bf R}_{\breve{\bm\theta}}[n,m']-{\bf H}[n,m'])\\
&=0,
\end{split}\end{equation}

and
\begin{equation}\begin{split}\label{sigma-sigma}
&\lim_{N\rightarrow\infty}\frac{1}{\sigma^2N}\sum_{n=1}^N\sum_{m=1}^M\left(\frac{{\bf R}_{\breve{\bm\theta}}[n,m]-{\bf H}[n,m]}{\sigma}\right)^2\\
&=\lim_{N\rightarrow\infty}\frac{1}{\sigma^4N}\sum_{n=1}^N\sum_{m=1}^M\left({\bf R}_{\breve{\bm\theta}}[n,m]-{\bf H}[n,m]\right)^2\\
&=\frac{1}{\sigma^4}\left[\frac{1}{2}\sum_{k=1}^K\sum_{m=1}^MA_{k,m}^2+\sigma^2_H\right],\\
&\sigma^2_H=\lim_{N\rightarrow\infty}\frac{1}{N}\sum_{n=1}^N\sum_{m=1}^M{\bf H}^2[n,m].
\end{split}\end{equation}

Define the normalization matrix ${\bf P}_N$ as
\begin{equation}
{\bf P}_N = \begin{bmatrix}
\frac{1}{N^{3/2}}{\bf I}_{K}&0\\
0&\frac{1}{N^{1/2}}{\bf I}_{2KM+1}
\end{bmatrix}.
\end{equation}
Using (\ref{omega-omega})-(\ref{sigma-sigma}), we have that
\begin{equation}\begin{split}\label{PJP}
&\lim_{N\rightarrow\infty}{\bf P}_N{\bf J}_1{\bf P}_N = \frac{1}{2\sigma^2}\\
&\left[
	\begin{array}{cccc}
	\tilde{\bf J}_{\omega\omega}& 0 &\tilde{\bf J}_{\omega\phi}&0\\
	0& {\bf I}_{KM} & 0&\tilde{\bf J}_{A\sigma}\\
	\tilde{\bf J}_{\omega\phi}^T& 0&\tilde{\bf J}_{\phi\phi}&0\\
	0&\tilde{\bf J}_{A\sigma}^T&0&
\frac{\sum_{k=1}^K\!\sum_{m=1}^M\!A_{k,m}^2+2\sigma^2_H}{\sigma^2}
	\end{array}
	\right],\\
&\tilde{\bf J}_{\omega\omega}=\begin{bmatrix}
\frac{1}{3}\sum_{m=1}^MA_{1,m}^2&&0\\
&\ddots&\\
0&&\frac{1}{3}\sum_{m=1}^MA_{K,m}^2
\end{bmatrix}\in\mathbb{R}^{K\times K},\\
&\tilde{\bf J}_{\omega\phi}\!=\!\!\begin{bmatrix}
\frac{A_{1,1}^2}{2}&\cdots&\frac{A_{1,M}^2}{2}&0&\cdots&\cdots&0\\
&\ddots&\ddots&\ddots&\ddots&\ddots&\\
0&\cdots&\cdots&0&\frac{A_{K,1}^2}{2}&\cdots&\frac{A_{K,M}^2}{2}
\end{bmatrix}\!\!\!\in\!\!\mathbb{R}^{K\times KM},\\
&\tilde{\bf J}_{A\sigma}=[-\frac{A_{1,1}}{\sigma},\dots, -\frac{A_{K,M}}{\sigma}]^T\in\mathbb{R}^{KM},
\end{split}\end{equation}
\begin{equation*}\begin{split}
&\tilde{\bf J}_{\phi\phi}=\begin{bmatrix}
A_{1,1}^2&&0\\
&\ddots&\\
0&&A_{K,M}^2
\end{bmatrix}\in\mathbb{R}^{KM\times KM},
\end{split}\end{equation*}
which implies, after straightforward calculations, that
\begin{equation}\begin{split}\label{detPJP}
\det(\lim_{N\rightarrow\infty}{\bf P}_N{\bf J}_1{\bf P}_N)&= \frac{1}{12^K}\left(\frac{1}{2\sigma^2}\right)^{K(1+2M)+1}\frac{2\sigma^2_H}{\sigma^2}\\
&\prod_{k=1}^K\left(\sum_{m=1}^MA_{k,m}^2\right)\prod_{k=1}^K\prod_{m=1}^MA_{k,m}^2.
\end{split}\end{equation}

Note that the asymptotic matrix in (\ref{PJP}) is nonsingular if and only if $\sigma_H^2\neq 0$ (i.e., the threshold is not zero). Combining (\ref{J-hatJ}) and (\ref{detPJP}), we get
\begin{equation}\begin{split}
\log\det(\hat{\bf J}({\bf Y},\hat{\breve{\bm\beta}})) &= \log\det({\bf P}_N^{-2})+\log\det({\bf P}_N{\bf J}_1{\bf P}_N)\\
& = (K(3+2M)+1)\log N+O(1).
\end{split}\end{equation}
Therefore, the extended 1bBIC metric is given by
\begin{equation}\begin{split}
-2\sum_{m=1}^{M}\sum_{n=1}^{N}\log\!\!\left[\!\Phi\!\left(\!{\bf Y}[n,m]\frac{{\bf R}_{\breve{\bm\theta}}[n,m]\!-\!{\bf H}[n,m]}{\sigma}\!\right)\!\right]\\
+K(3+2M)\log N.
\end{split}\end{equation}
where we only keep the terms that depend on $K$.


\ifCLASSOPTIONcaptionsoff
  \newpage
\fi



%
\bibliographystyle{IEEEtran}
\bibliography{journal}

\begin{thebibliography}{10}
\providecommand{\url}[1]{#1}
\csname url@samestyle\endcsname
\providecommand{\newblock}{\relax}
\providecommand{\bibinfo}[2]{#2}
\providecommand{\BIBentrySTDinterwordspacing}{\spaceskip=0pt\relax}
\providecommand{\BIBentryALTinterwordstretchfactor}{4}
\providecommand{\BIBentryALTinterwordspacing}{\spaceskip=\fontdimen2\font plus
\BIBentryALTinterwordstretchfactor\fontdimen3\font minus
  \fontdimen4\font\relax}
\providecommand{\BIBforeignlanguage}[2]{{%
\expandafter\ifx\csname l@#1\endcsname\relax
\typeout{** WARNING: IEEEtran.bst: No hyphenation pattern has been}%
\typeout{** loaded for the language `#1'. Using the pattern for}%
\typeout{** the default language instead.}%
\else
\language=\csname l@#1\endcsname
\fi
#2}}
\providecommand{\BIBdecl}{\relax}
\BIBdecl

\bibitem{CHND01}
C.~{Chen}, M.~B. {Higgins}, K.~{O'Neill}, and R.~{Detsch},
  ``Ultrawide-bandwidth fully-polarimetric ground penetrating radar
  classification of subsurface unexploded ordinance,'' \emph{IEEE Transactions
  on Geoscience and Remote Sensing}, vol.~39, no.~6, pp. 1221--1230, June 2001.

\bibitem{XN01}
X.~{Xu} and R.~M. {Narayanan}, ``{FOPEN} {SAR} imaging using {UWB}
  step-frequency and random noise waveforms,'' \emph{IEEE Transactions on
  Aerospace and Electronic Systems}, vol.~37, no.~4, pp. 1287--1300, Oct 2001.

\bibitem{YLML06}
A.~G. {Yarovoy}, L.~P. {Ligthart}, J.~{Matuzas}, and B.~{Levitas}, ``{UWB}
  radar for human being detection,'' \emph{IEEE Aerospace and Electronic
  Systems Magazine}, vol.~21, no.~3, pp. 10--14, March 2006.

\bibitem{SLL20}
X.~{Shang}, J.~{Liu}, and J.~{Li}, ``Multiple object localization and vital
  sign monitoring using {IR-UWB} {MIMO} radar,'' \emph{IEEE Transactions on
  Aerospace and Electronic Systems}, vol.~56, no.~6, pp. 4437--4450, 2020.

\bibitem{ZHLLW18}
B.~{Zhao}, L.~{Huang}, J.~{Li}, M.~{Liu}, and J.~{Wang}, ``Deceptive {SAR}
  jamming based on 1-bit sampling and time-varying thresholds,'' \emph{IEEE
  Journal of Selected Topics in Applied Earth Observations and Remote Sensing},
  vol.~11, no.~3, pp. 939--950, March 2018.

\bibitem{ZHB19}
B.~{Zhao}, L.~{Huang}, and W.~{Bao}, ``One-bit {SAR} imaging based on
  single-frequency thresholds,'' \emph{IEEE Transactions on Geoscience and
  Remote Sensing}, vol.~57, no.~9, pp. 7017--7032, 2019.

\bibitem{Xethru}
\BIBentryALTinterwordspacing
{Novelda AS}. (2017) Xethru: Single-chip radar sensors with sub-mm resolution.
  [Online]. Available: \url{https://www.xethru.com/}
\BIBentrySTDinterwordspacing

\bibitem{HWLLM07}
H.~A. {Hjortland}, D.~T. {Wisland}, T.~S. {Lande}, C.~{Limbodal}, and
  K.~{Meisal}, ``Thresholded samplers for {UWB} impulse radar,'' in \emph{2007
  IEEE International Symposium on Circuits and Systems}, May 2007, pp.
  1210--1213.

\bibitem{KL95}
T.~{Koutsoudis} and L.~A. {Lovas}, ``{RF} interference suppression in
  ultrawideband radar receivers,'' in \emph{Algorithms for Synthetic Aperture
  Radar Imagery II}, vol. 2487.\hskip 1em plus 0.5em minus 0.4em\relax
  International Society for Optics and Photonics, 1995, pp. 107--119.

\bibitem{C91}
D.~O. {Carhoun}, ``Adaptive nulling and spatial spectral estimation using an
  iterated principal components decomposition,'' in \emph{1991 International
  Conference on Acoustics, Speech, and Signal Processing}, April 1991, pp.
  3309--3312 vol.5.

\bibitem{SA93}
H.~{Subbaram} and K.~{Abend}, ``Interference suppression via orthogonal
  projections: a performance analysis,'' \emph{IEEE Transactions on Antennas
  and Propagation}, vol.~41, no.~9, pp. 1187--1194, Sep. 1993.

\bibitem{VSPH10}
V.~T. {Vu}, T.~K. {Sj\"{o}gren}, M.~I. {Pettersson}, and L.~{H{\aa}kasson},
  ``An approach to suppress {RFI} in ultrawideband low frequency {SAR},'' in
  \emph{2010 IEEE Radar Conference}, May 2010, pp. 1381--1385.

\bibitem{LUASF01}
X.~{Luo}, L.~M.~H. {Ulander}, J.~{Askne}, G.~{Smith}, and P.~O. {Frolind},
  ``{RFI} suppression in ultra-wideband {SAR} systems using {LMS} filters in
  frequency domain,'' \emph{Electronics Letters}, vol.~37, no.~4, pp. 241--243,
  Feb 2001.

\bibitem{MPM97}
T.~{Miller}, L.~{Potter}, and J.~{McCorkle}, ``{RFI} suppression for ultra
  wideband radar,'' \emph{IEEE Transactions on Aerospace and Electronic
  Systems}, vol.~33, no.~4, pp. 1142--1156, Oct 1997.

\bibitem{ZWXB07}
F.~{Zhou}, R.~{Wu}, M.~{Xing}, and Z.~{Bao}, ``Eigensubspace-based filtering
  with application in narrow-band interference suppression for {SAR},''
  \emph{IEEE Geoscience and Remote Sensing Letters}, vol.~4, no.~1, pp. 75--79,
  Jan 2007.

\bibitem{ZTBL13}
F.~{Zhou}, M.~{Tao}, X.~{Bai}, and J.~{Liu}, ``Narrow-band interference
  suppression for {SAR} based on independent component analysis,'' \emph{IEEE
  Transactions on Geoscience and Remote Sensing}, vol.~51, no.~10, pp.
  4952--4960, Oct 2013.

\bibitem{RZLNS19}
J.~{Ren}, T.~{Zhang}, J.~{Li}, L.~H. {Nguyen}, and P.~{Stoica}, ``{RFI}
  mitigation for {UWB} radar via hyperparameter-free sparse {SPICE} methods,''
  \emph{IEEE Transactions on Geoscience and Remote Sensing}, vol.~57, no.~6,
  pp. 3105--3118, June 2019.

\bibitem{RZLS19}
J.~{Ren}, T.~{Zhang}, J.~{Li}, and P.~{Stoica}, ``Sinusoidal parameter
  estimation from signed measurements via majorization-minimization based
  {RELAX},'' \emph{IEEE Transactions on Signal Processing}, vol.~67, no.~8, pp.
  2173--2186, April 2019.

\bibitem{ZRGL19}
T.~{Zhang}, J.~{Ren}, C.~{Gianelli}, and J.~{Li}, ``{RFI} mitigation for
  one-bit {UWB} radar systems,'' in \emph{2019 53rd Asilomar Conference on
  Signals, Systems, and Computers}, 2019, pp. 1545--1549.

\bibitem{BPCPE11}
S.~{Boyd}, N.~{Parikh}, E.~{Chu}, B.~{Peleato}, and J.~{Eckstein},
  ``Distributed optimization and statistical learning via the alternating
  direction method of multipliers,'' \emph{Foundations and Trends in Machine
  Learning}, vol.~3, no.~1, pp. 1--122, 2011.

\bibitem{LZLS18}
C.~{Li}, R.~{Zhang}, J.~{Li}, and P.~{Stoica}, ``Bayesian information criterion
  for signed measurements with application to sinusoidal signals,'' \emph{IEEE
  Signal Processing Letters}, vol.~25, no.~8, pp. 1251--1255, Aug 2018.

\bibitem{GXLS16}
C.~{Gianelli}, L.~{Xu}, J.~{Li}, and P.~{Stoica}, ``One-bit compressive
  sampling with time-varying thresholds: Maximum likelihood and the
  {C}ram\'{e}r-{R}ao bound,'' in \emph{2016 50th Asilomar Conference on
  Signals, Systems and Computers}, Nov 2016, pp. 399--403.

\bibitem{BV09}
\BIBentryALTinterwordspacing
S.~{Boyd} and L.~{Vandenberghe}, \emph{Convex optimization}.\hskip 1em plus
  0.5em minus 0.4em\relax Cambrige, UK: Cambridge university press, 2009.
  [Online]. Available:
  \url{http://stanford.edu/\~{}boyd/cvxbook/bv\_cvxbook.pdf}
\BIBentrySTDinterwordspacing

\bibitem{GXLS17}
C.~{Gianelli}, L.~{Xu}, J.~{Li}, and P.~{Stoica}, ``One-bit compressive
  sampling with time-varying thresholds for multiple sinusoids,'' in \emph{2017
  IEEE 7th International Workshop on Computational Advances in Multi-Sensor
  Adaptive Processing (CAMSAP)}, Dec 2017, pp. 1--5.

\bibitem{LS96}
J.~{Li} and P.~{Stoica}, ``Efficient mixed-spectrum estimation with
  applications to target feature extraction,'' \emph{IEEE Transactions on
  Signal Processing}, vol.~44, no.~2, pp. 281--295, Feb 1996.

\bibitem{Z69}
W.~I. {Zangwill}, \emph{Nonlinear programming: A unified approach}.\hskip 1em
  plus 0.5em minus 0.4em\relax Englewood Cliffs, N.J: Prentice-Hall, 1969.

\bibitem{HL04}
D.~{Hunter} and K.~{Lange}, ``A tutorial on {MM} algorithms,'' \emph{The
  American Statistician}, vol.~58, no.~1, pp. 30--37, 2004.

\bibitem{SS04}
P.~{Stoica} and Y.~{Selen}, ``Cyclic minimizers, majorization techniques, and
  the expectation-maximization algorithm: a refresher,'' \emph{IEEE Signal
  Processing Magazine}, vol.~21, no.~1, pp. 112--114, Jan 2004.

\bibitem{LZS97}
J.~{Li}, D.~{Zheng}, and P.~{Stoica}, ``Angle and waveform estimation via
  {RELAX},'' \emph{IEEE Transactions on Aerospace and Electronic Systems},
  vol.~33, no.~3, pp. 1077--1087, July 1997.

\bibitem{BB08}
P.~T. {Boufounos} and R.~G. {Baraniuk}, ``1-bit compressive sensing,'' in
  \emph{2008 42nd Annual Conference on Information Sciences and Systems}, March
  2008, pp. 16--21.

\bibitem{NTD14}
L.~H. {Nguyen}, T.~{Tran}, and T.~{Do}, ``Sparse models and sparse recovery for
  ultra-wideband {SAR} applications,'' \emph{IEEE Transactions on Aerospace and
  Electronic Systems}, vol.~50, no.~2, pp. 940--958, April 2014.

\bibitem{NT16}
L.~H. {Nguyen} and T.~D. {Tran}, ``Efficient and robust {RFI} extraction via
  sparse recovery,'' \emph{IEEE Journal of Selected Topics in Applied Earth
  Observations and Remote Sensing}, vol.~9, no.~6, pp. 2104--2117, June 2016.

\bibitem{RNKWS07}
M.~{Ressler}, L.~{Nguyen}, F.~{Koenig}, D.~{Wong}, and G.~{Smith}, ``{The Army
  Research Laboratory ({ARL}) synchronous impulse reconstruction ({SIRE})
  forward-looking radar},'' in \emph{Unmanned Systems Technology IX}, G.~R.
  Gerhart, D.~W. Gage, and C.~M. Shoemaker, Eds., vol. 6561, International
  Society for Optics and Photonics.\hskip 1em plus 0.5em minus 0.4em\relax
  SPIE, 2007, pp. 35 -- 46.

\bibitem{SS04-2}
P.~{Stoica} and Y.~{Selen}, ``Model-order selection: a review of information
  criterion rules,'' \emph{IEEE Signal Processing Magazine}, vol.~21, no.~4,
  pp. 36--47, July 2004.

\bibitem{SMFS98}
P.~{Stoica}, R.~L. {Moses}, B.~{Friedlander}, and T.~{Soderstrom}, ``Maximum
  likelihood estimation of the parameters of multiple sinusoids from noisy
  measurements,'' \emph{IEEE Transactions on Acoustics, Speech, and Signal
  Processing}, vol.~37, no.~3, pp. 378--392, March 1989.

\end{thebibliography}
%






\end{document}